\documentclass[3p,times]{elsarticle}
\usepackage{mathtools}
\usepackage{xcolor}
\usepackage[numbers]{natbib} 
\usepackage[utf8]{inputenc}
\usepackage[english]{babel}
\usepackage{todonotes} % for comments
\usepackage[colorlinks,allcolors=blue]{hyperref}
\usepackage{caption}

\usepackage{subcaption,siunitx,booktabs}
\usepackage{colortbl}
\usepackage{amsmath,amsfonts,amsthm,bm}
\usepackage{float}

\usepackage{multirow}
\usepackage{diagbox}
\usepackage{tabularx}
\newcolumntype{Y}{>{\raggedright\arraybackslash}X}
\usepackage{setspace}
\usepackage{threeparttable}

\begin{document}
\doublespacing
\begin{frontmatter}

\title{When market boundaries weaken: Network reconfiguration and regime-dependent cross-asset spillovers}

\author[1]{Ruixue Jing}
\author[1,2]{Luis E. C. Rocha}
\address[1]{Department of Economics, Ghent University, Ghent, 9000, Belgium}
\address[2]{Department of Physics and Astronomy, Ghent University, Ghent, 9000, Belgium}

\begin{abstract}
Cryptocurrencies are increasingly adopted as investment assets, making their interactions with traditional financial markets central to cross-asset diversification and systemic risk. This paper studies the integration of cryptocurrencies, fiat currencies, and S\&P500 equities using a balanced panel of 381 assets from October 2017 to February 2024. We combine rolling correlation networks, community structure, market-specific and system-wide Turbulence Indices, and VAR-based connectedness analysis to examine how market stress, network structure, and shock transmission vary across financial regimes. The results show that cross-asset integration is episodic. In calm periods, the three asset classes remain relatively segmented, whereas under stress, local clustering increases, modular separation weakens, and communities become more compositionally mixed across asset classes. Connectedness analysis further shows that regime shifts alter the structure of transmission rather than simply increasing spillover magnitudes. In high-turbulence states, fiat-market turbulence becomes the dominant propagation channel, while network clustering and modularity play a greater role in transmitting forecast uncertainty. These findings support the interpretation of network structure as an emergent, state-dependent transmission layer rather than a persistent exogenous driver of turbulence. The results highlight the need for regime-aware risk monitoring, since full-sample connectedness estimates can understate the cross-asset coupling that emerges precisely when diversification benefits are most fragile.
\end{abstract}


\end{frontmatter}

\section{Introduction}

Cross-asset diversification depends on whether markets remain imperfectly integrated during stress. A recurring empirical regularity is that correlations and volatility spillovers rise during episodes of systemic stress, compressing the diversification benefits that investors and risk managers ordinarily expect from positions across distinct markets~\citep{billio2012econometric,adrian2011covar,acharya2017measuring,forbes2002no,longin2001extreme,bekaert2003market,bekaert2014global,dungey2005empirical}. This regularity is well documented, but its transmission structure remains less well understood. Standard explanations emphasise common-factor exposure or direct transmission through financial intermediaries~\citep{forbes2002no,dungey2005empirical}, yet they do not fully explain why cross-asset coupling varies between calm and turbulent states, or why the identity of markets that transmit and absorb stress may change when market conditions deteriorate. These questions have become more important since 2017, as cryptocurrencies have become more actively traded alongside equities and currencies during a period marked by global macro-financial stress and major cryptocurrency-specific disruptions. Understanding cross-asset integration, therefore, requires asking when diversification benefits weaken, which markets transmit stress, and how the dependence structure linking asset classes vary across regimes.

Existing finance research has mainly addressed this problem using time-varying dependence, volatility spillover, and connectedness models. Multivariate GARCH and Dynamic Conditional Correlation frameworks provide standard tools for modelling time-varying conditional volatility and cross-market correlations~\citep{bollerslev1990modelling,engle1995multivariate,engle2002dynamic,tse2002multivariate,aielli2013dynamic}. However, crisis episodes also reveal that dependence can be nonlinear and state-contingent. Correlations may increase disproportionately during extreme market conditions, so average or full-sample correlations can underestimate dependence in turbulent states~\citep{longin2001extreme,patton2006modelling,embrechts2002correlation}. This motivates a regime-dependent view of cross-asset dependence, in which calm and turbulent periods may not simply differ in the magnitude of correlations, but may correspond to different transmission environments. Complementary literature examines dynamic connectedness using multivariate time-series systems. VAR-based connectedness models instead focus on predictive transmission and forecast-error variance spillovers~\citep{sims1980macroeconomics,kilian2006new,granger1969investigating}. The Diebold--Yilmaz connectedness framework is used to measure total, directional, and net spillovers from forecast error variance decompositions~\citep{diebold2009measuring,diebold2012better,diebold2014network,antonakakis2020refined,gabauer2020volatility}. Generalised variance decompositions are particularly useful because they reduce sensitivity to variable ordering~\citep{pesaran1998generalized,koop1996impulse}. This literature shifts the focus from static association to predictive transmission. However, full-sample and rolling estimates may still hide whether low- and high-stress states correspond to different transmission structures.

This issue is especially relevant for cryptocurrencies, whose diversification and spillover roles remain state-dependent. Existing evidence suggests that crypto--traditional market spillovers are asymmetric and sensitive to global risk conditions~\citep{koutmos2018return,kurka2019cryptocurrencies,bouri2018spillovers,baur2018bitcoin,klein2018bitcoin,ji2019dynamic,dyhrberg2016bitcoin}. Several studies find that traditional markets, volatility expectations, and macro-financial uncertainty often transmit more strongly into cryptocurrencies than cryptocurrencies transmit back into traditional markets. This evidence suggests that cryptocurrencies are neither fully isolated from traditional markets nor necessarily dominant systemic transmitters. Their role is conditional on the market environment, making the distinction between average integration and stress-state integration important for portfolio construction and systemic risk monitoring. The key issue is therefore not whether cryptocurrencies are connected to traditional markets on average, but whether their role changes when the whole cross-asset system enters a stressed state.

Network methods provide a complementary way to study the structure of financial interdependence~\citep{mantegna1999hierarchical,onnela2003dynamics,tumminello2005tool,aste2005correlation,newman2003mixing}. In correlation-based financial networks, assets are represented as nodes and pairwise dependence defines links or distances. Correlation-based networks summarise not only the average level of comovement, but also the structure of market organisation. Measures such as clustering, modularity, and community structure indicate whether dependence is locally concentrated, globally segmented, or mixed across asset classes. This matters because stress may reorganise dependence in ways that average correlations do not capture. Prior network studies show that crises are associated with stronger comovement, shorter effective distances, denser local structures, and changes in community organisation~\citep{onnela2003dynamics,tumminello2005tool,aste2005correlation}. However, interpreting these network measures requires caution. Correlation networks describe comovement rather than structural causal maps. Their structure emerges from market behaviour and should therefore be interpreted as a state-dependent pattern of dependence rather than as an exogenous driver of returns.

Although the connectedness and network literatures are closely related, they are often used separately. Connectedness models quantify directional spillovers and transmitter--receiver roles, but usually say less about how dependence structure changes over time. Network studies describe market structure, but they are often not embedded in the dynamic systems used to estimate shock transmission. This separation is limiting if the network structure becomes more closely linked to short-run propagation during stress rather than merely summarising instantaneous comovement. Financial-network theory shows that the pattern of interconnections can affect whether shocks are diversified, absorbed, or amplified~\citep{acemoglu2015systemic,elliott2014financial}. This motivates testing whether network structure is only a contemporaneous description of market organisation, or whether changes in clustering, modularity, and cross-class mixing are also linked to subsequent forecast-uncertainty transmission when the system is already turbulent.

Currency markets are especially relevant in this context. The global financial cycle literature links dollar conditions and exchange-rate movements to funding liquidity and cross-border capital flows~\citep{miranda2020us,avdjiev2019dollar,brunnermeier2009market}. During stress, currency-market turbulence may therefore reflect more than exchange-rate volatility alone. It can summarise shifts in global funding conditions and risk appetite that may affect both traditional and digital asset markets. This suggests that fiat turbulence may have a distinct role in cross-asset transmission rather than serving only as another market-specific stress measure. This role may become especially important when global risk conditions deteriorate, and dollar-denominated funding and portfolio rebalancing channels become more relevant. State dependence also requires a regime definition based on system-wide stress rather than on a single market. A threshold VAR provides a tractable way to examine whether propagation differs between low- and high-stress conditions~\citep{tong2012threshold,tsay1998testing,hansen2000sample}. A Mahalanobis-based turbulence index is useful because it identifies unusual joint return movements while accounting for covariance structure~\citep{kritzman2010skulls,de2000mahalanobis}. A system-wide turbulence measure is therefore preferable to a single-market trigger because it identifies unusual joint configurations of the full cross-asset system.

This paper contributes to the finance literature on cross-asset connectedness by showing how diversification fragility, dependence structure, and transmitter-receiver roles jointly change across turbulence regimes. It studies cryptocurrencies, fiat currencies, and S\&P500 equities in a balanced design, allowing turbulence, dependence structure, and spillover dynamics to be compared across asset classes. Then, it incorporates time-varying network indicators into a VAR-based connectedness system, so that clustering, modularity, and cross-class community diversity are analysed jointly with market turbulence and macro-financial conditions. Finally, it examines whether the role of network structure differs across turbulence regimes, distinguishing ordinary cross-asset comovement from stress-state transmission. This distinction separates calendar-time variation in connectedness from state dependence, showing whether the same variables play different roles when the system moves from calm to stressed conditions. The proposed framework, therefore, assesses when network structure mainly describes market organisation and when it becomes part of the short-run spillover environment. The results show that cross-asset integration is episodic. In calm periods, cryptocurrencies, fiat currencies, and equities remain partly segmented, whereas stress periods are associated with stronger local clustering, weaker modular separation, and greater cross-class mixing. The regime-dependent connectedness results show that fiat turbulence becomes the dominant transmitter in high-turbulence states, whereas network clustering and modularity become more involved in transmitting forecast uncertainty.

\section{Materials and Methods}

\subsection{Market turbulence and regime classification}
\label{sec:methods_turbulence}

Let \(P_{i,t}\) denote the price of asset \(i\) at time \(t\). Daily log returns are computed as
\begin{equation}
r_{i,t} = \ln \left( \frac{P_{i,t}}{P_{i,t-1}} \right).
\label{eq:log_return}
\end{equation}
These log returns are used to compute both the Turbulence Indices and the rolling correlation networks.

Before computing the Turbulence Index~\citep{kritzman2010skulls}, each asset's daily log return series is standardised as
\begin{equation}
z_{i,t} = \frac{r_{i,t}-\bar r_i}{s_i},
\label{eq:standardised_return}
\end{equation}
where \(\bar r_i\) and \(s_i\) are the sample mean and standard deviation of the daily log returns of asset \(i\). This standardisation is performed over the full sample, while the rolling mean vector and covariance matrix used in the Turbulence Index are estimated only from the previous $\Delta_{\mathrm{TI}}$ observations.

For a given market \(m\), let \(\mathbf{z}_{m,t}\in\mathbb{R}^{N_m}\) denote the vector of standardised daily log returns for the \(N_m\) assets in market \(m\). The market-specific Turbulence Index is computed as a regularised squared Mahalanobis distance~\citep{de2000mahalanobis} between the current standardised return vector and its recent rolling distribution:
\begin{equation}
\mathrm{TI}_{m,t} = (\mathbf{z}_{m,t}-\boldsymbol{\mu}_{m,t})^{\top} \left(\boldsymbol{\Sigma}^{\mathrm{reg}}_{m,t}\right)^{+} (\mathbf{z}_{m,t}-\boldsymbol{\mu}_{m,t}),
\label{eq:market_ti}
\end{equation}
where \(\boldsymbol{\mu}_{m,t}\) and \(\boldsymbol{\Sigma}^{\mathrm{reg}}_{m,t}\) are estimated from the previous \(\Delta_{\mathrm{TI}}\) daily observations. The superscript \(+\) denotes the Moore--Penrose pseudoinverse, which is used as a numerical safeguard in the high-dimensional rolling-window setting. 

The covariance matrix is estimated using the Ledoit--Wolf shrinkage estimator~\citep{ledoit2004well}. This estimator stabilises the high-dimensional rolling covariance estimate by shrinking the empirical covariance matrix toward a scaled identity target, reducing sensitivity to sampling noise when the number of observations is small relative to the number of assets. We then add a small diagonal perturbation:
\begin{equation}
\boldsymbol{\Sigma}^{\mathrm{reg}}_{m,t} = \boldsymbol{\Sigma}^{\mathrm{LW}}_{m,t} + \lambda \mathbf{I},
\label{eq:cov_regularisation}
\end{equation}
where \(\boldsymbol{\Sigma}^{\mathrm{LW}}_{m,t}\) denotes the Ledoit--Wolf covariance estimate, \(\lambda=10^{-5}\), and \(\mathbf{I}\) is the \(N_m \times N_m\) identity matrix. The diagonal perturbation is included only for numerical regularisation, while the main stabilisation comes from the Ledoit--Wolf shrinkage estimator.

For the full cross-asset system, let \(\mathbf{z}_{t}\in\mathbb{R}^{N}\) collect the standardised daily log returns of all assets across the asset classes. The system-wide Turbulence Index is:
\begin{equation}
\mathrm{TI}_{t} = (\mathbf{z}_{t}-\boldsymbol{\mu}_{t})^{\top} \left(\boldsymbol{\Sigma}^{\mathrm{reg}}_{t}\right)^{+} (\mathbf{z}_{t}-\boldsymbol{\mu}_{t}).
\label{eq:system_ti}
\end{equation}
This system-wide index is used for regime classification, while the three market-specific Turbulence indices enter the VAR system as endogenous variables.

\subsection{Dynamic connectedness and regime-dependent transmission}
\label{sec:methods_dynamic_transmission}

We use a vector autoregression (VAR)~\citep{sims1980macroeconomics} to study how changes in market turbulence, network structure, and macro-financial conditions propagate through the system over time. The same model is then used to compute connectedness measures and to compare how spillover structure changes across turbulence regimes.

Before estimating the VAR, each endogenous series is transformed into a standardised first difference. For any variable \(X_t\), we define
\begin{equation}
\widetilde{X}_t = \frac{\Delta X_t-\overline{\Delta X}}{s_{\Delta X}}, 
\qquad
\Delta X_t = X_t-X_{t-1},
\label{eq:standardised_difference}
\end{equation}
where \(\overline{\Delta X}\) and \(s_{\Delta X}\) are the sample mean and sample standard deviation of \(\Delta X_t\). This transformation places all endogenous variables on a comparable scale and focuses the VAR on changes rather than persistent level differences. The endogenous vector $\mathbf{Y}_t$ contains the standardised first differences of $\mathrm{TI}_{\mathrm{s\&p500},t}$, $\mathrm{TI}_{\mathrm{fiat},t}$, $\mathrm{TI}_{\mathrm{crypto},t}$, $\langle cc\rangle_t$, $Q_t$, $\langle D\rangle_t$, $\mathrm{VIX}_t$, $\mathrm{USDX}_t$, and $\mathrm{EPU}_t$.

The baseline VAR with \(p\) lags is specified as
\begin{equation}
\mathbf{Y}_t = \boldsymbol{\alpha} + \sum_{\ell=1}^{p} \boldsymbol{\Phi}_{\ell}\mathbf{Y}_{t-\ell} + \boldsymbol{u}_t,
\label{eq:var}
\end{equation}
where \(\boldsymbol{\alpha}\) is a vector of intercepts, \(\boldsymbol{\Phi}_{\ell}\) is the coefficient matrix for lag \(\ell\), and \(\boldsymbol{u}_t\) is the vector of model residuals. Their covariance matrix is denoted by
\begin{equation}
\boldsymbol{\Sigma}_{u} = \mathbb{E}(\boldsymbol{u}_t\boldsymbol{u}_t^{\top}).
\label{eq:residual_covariance}
\end{equation}
We compare candidate lag lengths using standard information criteria, including AIC~\citep{akaike1974new}, BIC~\citep{schwarz1978estimating}, HQIC~\citep{hannan1979determination}, and FPE~\citep{akaike1969fitting}. After estimation, VAR stability is verified by confirming that all eigenvalues of the companion matrix lie inside the unit circle, meaning that shocks to the system eventually dissipate rather than grow indefinitely.

To measure the magnitude of transmission, we use the Diebold--Yilmaz connectedness framework~\citep{diebold2012better,diebold2014network} based on the generalised forecast error variance decomposition (GFEVD)~\citep{pesaran1998generalized}. The idea is to decompose the forecast uncertainty of each variable into parts attributable to itself and parts attributable to other variables. For a stable estimated VAR, the model can be written in moving-average form for the centred process as 
\begin{equation}
\mathbf{Y}_t - \boldsymbol{\mu}_{Y} = \sum_{h=0}^{\infty} \boldsymbol{\Psi}_{h}\boldsymbol{u}_{t-h},
\label{eq:ma_representation}
\end{equation}
where \(\boldsymbol{\mu}_{Y}\) denotes the unconditional mean of \(\mathbf{Y}_t\), and \(\boldsymbol{\Psi}_{h}\) describes how a model residual propagates after \(h\) days. 
The \(H\)-day generalised variance share from variable \(j\) to variable \(i\) is
\begin{equation}
\theta^{(H)}_{ij} = \frac{ \sigma_{u,jj}^{-1} \sum_{h=0}^{H-1} \left( \mathbf{e}_i^{\top} \boldsymbol{\Psi}_h \boldsymbol{\Sigma}_{u} \mathbf{e}_j \right)^2 }
{ \sum_{h=0}^{H-1} \mathbf{e}_i^{\top} \boldsymbol{\Psi}_h \boldsymbol{\Sigma}_{u} \boldsymbol{\Psi}_h^{\top} \mathbf{e}_i },
\label{eq:gfevd}
\end{equation}
where \(\mathbf{e}_i\) is a selection vector for variable \(i\), and \(\sigma_{u,jj}\) is the \(j\)-th diagonal element of \(\boldsymbol{\Sigma}_{u}\). We use the generalised decomposition because it is invariant to the ordering of variables in the VAR.

Because the generalised variance shares do not necessarily sum to one across each row, we normalise them as
\begin{equation}
\widetilde{\theta}^{(H)}_{ij} = \frac{\theta^{(H)}_{ij}} {\sum_{j=1}^{K}\theta^{(H)}_{ij}},
\qquad
\sum_{j=1}^{K} \widetilde{\theta}^{(H)}_{ij} = 1,
\label{eq:gfevd_normalised}
\end{equation}
where \(K\) is the number of variables in \(\mathbf{Y}_t\). All connectedness measures are reported in percentage terms. 
For each variable \(i\), spillovers received from the rest of the system are defined as
\begin{equation}
\mathrm{From}^{(H)}_i = \sum_{j\neq i} \widetilde{\theta}^{(H)}_{ij},
\label{eq:from}
\end{equation}
while spillovers transmitted from variable \(i\) to the rest of the system are defined as
\begin{equation}
\mathrm{To}^{(H)}_i = \sum_{j\neq i} \widetilde{\theta}^{(H)}_{ji}.
\label{eq:to}
\end{equation}
The net spillover position is
\begin{equation}
\mathrm{Net}^{(H)}_i = \mathrm{To}^{(H)}_i - \mathrm{From}^{(H)}_i.
\label{eq:net}
\end{equation}
A positive value means that variable \(i\) transmits more forecast uncertainty than it receives, whereas a negative value means that it receives more than it transmits. 
The total spillover index is
\begin{equation}
\mathrm{TSI}^{(H)} = \frac{1}{K} \sum_{i=1}^{K} \sum_{j\neq i} \widetilde{\theta}^{(H)}_{ij}.
\label{eq:tsi}
\end{equation}
This index summarises the average share of forecast uncertainty that comes from cross-variable spillovers rather than from a variable's own residuals.

To examine how connectedness changes over time, we apply the VAR-GFEVD calculation in rolling windows. For each rolling window, we estimate the VAR and compute the corresponding total, directional, and net connectedness measures. This produces time-varying series $\mathrm{TSI}^{(H)}_t, \mathrm{To}^{(H)}_{i,t}, \mathrm{From}^{(H)}_{i,t}, \mathrm{Net}^{(H)}_{i,t}$. The rolling analysis is used to assess time variation in aggregate connectedness and net transmitter--receiver roles.

The rolling analysis describes time variation along the calendar dimension, but it does not directly separate calm and elevated-turbulence states. To test whether the system behaves differently across turbulence states, we estimate a two-regime threshold VAR~\citep{tong2012threshold,tsay1998testing}. The regime is defined directly from the system-wide Turbulence Index in levels, \(\mathrm{TI}_t\). We use a one-day delay, so the regime for day \(t\) depends on the previous day's system-wide turbulence:
\begin{equation}
s_t = \begin{cases}
\mathrm{Low}, & \mathrm{TI}_{t-1}\leq \gamma,\\
\mathrm{High}, & \mathrm{TI}_{t-1}>\gamma.
\end{cases}
\label{eq:tvar_state}
\end{equation}
Here, \(\gamma\) is the estimated turbulence threshold. The one-day delay avoids using the current day's system-wide turbulence to define the same day's regime.

The threshold VAR is then
\begin{equation}
\mathbf{Y}_t = \boldsymbol{\alpha}^{(s_t)} + \sum_{\ell=1}^{p} \boldsymbol{\Phi}^{(s_t)}_{\ell}\mathbf{Y}_{t-\ell} + \boldsymbol{u}^{(s_t)}_t,
\label{eq:tvar}
\end{equation}
where the intercepts, lag coefficients, and residual covariance matrix are allowed to differ between the low- and high-turbulence regimes. The threshold \(\gamma\) is selected by grid search. Candidate thresholds are restricted to the range between the 10th and 90th percentiles of \(\mathrm{TI}_{t-1}\), so that both regimes contain enough observations. For each candidate threshold, we estimate the two-regime VAR and select the threshold that minimises the Bayesian Information Criterion (BIC). We then check the stability of the VAR separately in each regime.

For each regime \(s\in\{\mathrm{Low},\mathrm{High}\}\), we compute regime-specific connectedness measures: $\widetilde{\theta}^{(H,s)}_{ij}, \mathrm{To}^{(H,s)}_i$, $\mathrm{From}^{(H,s)}_i$, $\mathrm{Net}^{(H,s)}_i$, $\mathrm{TSI}^{(H,s)}$. To summarise how transmission roles change under stress, we define
\begin{equation}
\Delta \mathrm{Net}^{(H)}_i = \mathrm{Net}^{(H,\mathrm{High})}_i - \mathrm{Net}^{(H,\mathrm{Low})}_i.
\label{eq:delta_net}
\end{equation}
A positive \(\Delta \mathrm{Net}^{(H)}_i\) means that variable \(i\) becomes more transmissive in the high-turbulence regime. A negative value means that it becomes more absorptive.

We use generalised impulse response functions (GIRFs) to visualise the dynamic adjustment behind the connectedness measures~\citep{pesaran1998generalized}. A GIRF traces the expected response of one variable after an unexpected one-standard-deviation change in another variable. Because we use the generalised version, the responses do not depend on the ordering of variables in the VAR. The response of variable \(q\) to a shock in variable \(v\) at horizon \(h\) is
\begin{equation}
\mathrm{GIRF}_{q\leftarrow v}(h) = \frac{ \left[ \boldsymbol{\Psi}_h \boldsymbol{\Sigma}_{u} \right]_{qv} }{ \sqrt{\sigma_{u,vv}} }.
\label{eq:girf}
\end{equation}
For the threshold VAR, the GIRFs are computed separately in the low- and high-turbulence regimes using the corresponding regime-specific coefficients and covariance matrices. Uncertainty is assessed using bootstrap confidence intervals. For the linear VAR, we resample model residuals, simulate bootstrap series from the estimated model, re-estimate the VAR, and recompute the GIRFs. For the threshold VAR, residuals are resampled within each regime, and responses are recomputed using the corresponding regime-specific coefficient matrices. Confidence bands are reported at the 90\% level. The interpretation focuses on the size, sign, persistence, and regime contrast of the responses.

Finally, we summarise transmission across three groups of variables for market turbulence, network structure, and macro-financial conditions. Let \(\mathcal{G}_a\) and \(\mathcal{G}_b\) denote two such groups. The spillover from group \(b\) to group \(a\) is
\begin{equation}
\Theta^{(H)}_{a\leftarrow b} = \frac{1}{|\mathcal{G}_a|} \sum_{i\in \mathcal{G}_a} \sum_{j\in \mathcal{G}_b} \widetilde{\theta}^{(H)}_{ij}.
\label{eq:block_spillover}
\end{equation}
Group-level received, transmitted, and net connectedness are then
\begin{equation}
\mathrm{From}^{(H)}_a = \sum_{b\neq a} \Theta^{(H)}_{a\leftarrow b},
\qquad
\mathrm{To}^{(H)}_a = \sum_{b\neq a} \Theta^{(H)}_{b\leftarrow a},
\label{eq:block_to_from}
\end{equation}
and
\begin{equation}
\mathrm{Net}^{(H)}_a = \mathrm{To}^{(H)}_a - \mathrm{From}^{(H)}_a.
\label{eq:block_net}
\end{equation}
These block measures show whether transmission mainly originates in market turbulence, macro-financial conditions, or the network structure of cross-asset dependence.

\subsection{Correlation networks and structural indicators}
\label{sec:correlation_networks}

We represent the cross-asset dependence structure at each time \(t\) as an undirected weighted network \(G_t=(V,\mathbf{A}_t,\mathbf{W}_t)\), where \(V\) is the set of assets and \(N=|V|\) is the number of nodes. The adjacency matrix \(\mathbf{A}_t=\{a_{ij,t}\}\) indicates whether two assets are connected, with \(a_{ij,t}=1\) if assets \(i\) and \(j\) are linked at time \(t\), and \(a_{ij,t}=0\) otherwise. The weighted adjacency matrix \(\mathbf{W}_t=\{w_{ij,t}\}\) contains the strength of each positive comovement, with \(w_{ij,t}>0\) for connected pairs and \(w_{ij,t}=0\) otherwise.

The number of edges and the total edge weight are defined as
\begin{equation}
E_t=\sum_{i<j}a_{ij,t},
\qquad
M_t=\sum_{i<j}w_{ij,t}=\frac{1}{2}\sum_{i,j}w_{ij,t}.
\label{eq:network_size_weight}
\end{equation}

All network quantities are computed in rolling windows. We denote the network estimation window by \(L_{\mathrm{net}}\) and set \(L_{\mathrm{net}}=30\) days, following the sampling-window choice used for short-run financial network estimation~\citep{rocha2017sampling}. Thus, \(\mathcal{R}_t=\{t-L_{\mathrm{net}}+1,\ldots,t\}\) is the return window used to estimate \(\rho_{ij,t}\), the adjacency matrix, and the correlation weights at time \(t\). 

Using the daily log returns defined in Eq.~\eqref{eq:log_return}, the rolling Pearson correlation between assets $i$ and $j$ is
\begin{equation}
\rho_{ij,t} = \frac{ \sum_{\tau\in \mathcal{R}_t} (r_{i,\tau}-\bar r_{i,t}) (r_{j,\tau}-\bar r_{j,t}) }{ \sqrt{ \sum_{\tau\in \mathcal{R}_t} (r_{i,\tau}-\bar r_{i,t})^2 } \sqrt{ \sum_{\tau\in \mathcal{R}_t} (r_{j,\tau}-\bar r_{j,t})^2 }},
\label{eq:correlation}
\end{equation}
where \(\bar r_{i,t}\) is the mean return of asset \(i\) within \(\mathcal{R}_t\).

The baseline network retains only positive return comovements. Specifically, we apply the correlation threshold $\theta_{\rho}=0$, and define the adjacency matrix as
\begin{equation}
a_{ij,t} = \mathbb{I}(\rho_{ij,t}>\theta_{\rho}), \qquad a_{ii,t}=0.
\label{eq:adjacency_correlation}
\end{equation}
The corresponding edge weight is $w_{ij,t} = \rho_{ij,t}a_{ij,t}$. Thus, connected pairs have positive correlation weights, while non-positive correlations are excluded from the network. This focuses the analysis on positive comovement, interpreted as assets moving together within the rolling window.

The baseline network measures are computed from this positive-correlation weighted network. As a robustness check, we also compute the corresponding correlation-distance representation $d_{ij,t}=\sqrt{2(1-\rho_{ij,t})}$, where shorter distances indicate stronger positive comovement~\citep{mantegna1999hierarchical} (see SI).

We summarise the network using three structural indicators that correspond to different levels of market organisation. Local cohesion is measured by the average weighted clustering coefficient. 
The degree of node \(i\) is
\begin{equation}
k_{i,t} = \sum_{j\neq i}a_{ij,t},
\label{eq:degree}
\end{equation}
and its weighted degree, or strength, is
\begin{equation}
\kappa_{i,t} = \sum_{j\neq i}w_{ij,t}.
\label{eq:weighted_degree}
\end{equation}
We use the Barrat weighted clustering coefficient:
\begin{equation}
cc_{i,t} = \frac{1}{\kappa_{i,t}(k_{i,t}-1)} \sum_{j,h} \frac{w_{ij,t}+w_{ih,t}}{2} a_{ij,t}a_{ih,t}a_{jh,t},
\label{eq:weighted_clustering}
\end{equation}
for nodes with \(k_{i,t}>1\). Here, \(k_{i,t}\) counts the number of neighbours of node \(i\), whereas \(\kappa_{i,t}\) measures the total weight of its connections. 
The network-level clustering coefficient is
\begin{equation}
\langle cc\rangle_t = \frac{1}{N_t^{(cc)}} \sum_{i:k_{i,t}>1} cc_{i,t},
\label{eq:average_clustering}
\end{equation}
where \(N_t^{(cc)}\) is the number of nodes with degree greater than one. A higher \(\langle cc\rangle_t\) indicates that assets are embedded in tighter local triangles of comovement, reflecting stronger local cohesion in the cross-asset dependence network.

Global segmentation is measured by modularity. For a partition of the network into communities, let \(g_{i,t}\) denote the community assignment of asset \(i\). The weighted modularity is
\begin{equation}
Q_t = \frac{1}{2M_t} \sum_{i,j} \left[ w_{ij,t} - \frac{\kappa_{i,t}\kappa_{j,t}}{2M_t} \right] \delta(g_{i,t},g_{j,t}),
\label{eq:modularity}
\end{equation}
where the Kronecker delta \(\delta(g_{i,t},g_{j,t})=1\) if assets \(i\) and \(j\) belong to the same community and zero otherwise. Higher \(Q_t\) indicates stronger separation into internally cohesive but mutually distinct communities. We use the Louvain algorithm~\citep{blondel2008fast} to obtain modular partitions by maximising \(Q_t\).

Because modularity maximisation is stochastic and can be sensitive to sampling variability~\citep{fortunato2007resolution,good2010performance}, we use a consensus clustering procedure to identify robust communities. This follows the consensus-clustering methodology used in previous financial analysis and adapts it to our cross-asset setting~\citep{jing2025optimising}.

For each reference time \(t\), we apply Louvain to \(n_s\) adjacent rolling windows indexed by \(b=1,\ldots,n_s\), each shifted by one day relative to the previous window. The co-membership frequency of assets \(i\) and \(j\) is
\begin{equation}
\pi_{ij,t} = \frac{1}{n_s} \sum_{b=1}^{n_s} \mathbb{I} \left( g^{(b)}_{i,t} = g^{(b)}_{j,t} \right).
\label{eq:consensus_similarity}
\end{equation}
The matrix \(\boldsymbol{\Pi}_t=\{\pi_{ij,t}\}\) records how often each pair of assets is assigned to the same community across adjacent windows. We retain pairs with
\begin{equation}
\pi_{ij,t}\geq \theta_{\pi}, \qquad \theta_{\pi}=0.5.
\label{eq:consensus_threshold}
\end{equation}
The connected components of this thresholded co-membership network define robust clusters. Thus, two assets are treated as belonging to the same robust cluster only if they co-occur in the same detected community in at least half of the repeated realisations. This procedure reduces dependence on a single Louvain realisation and focuses the analysis on persistent clustering.

Cross-class mixing is measured using Simpson's Diversity Index~\citep{simpson1949measurement}. For a robust cluster \(C\), let \(n_{C,t}\) denote the number of assets in the cluster, and let \(n_{x,C,t}\) denote the number of assets belonging to asset class \(x\). The cluster-level diversity index is
\begin{equation}
D_{C,t} = 1 - \frac{ \sum_{x=1}^{X} n_{x,C,t}(n_{x,C,t}-1) }{ n_{C,t}(n_{C,t}-1) },
\label{eq:simpson}
\end{equation}
where \(X\) is the number of asset classes. Equivalently, for sufficiently large clusters, if \(p_{x,C,t}=n_{x,C,t}/n_{C,t}\), then
\begin{equation}
D_{C,t} \approx 1-\sum_{x=1}^{X}p_{x,C,t}^{2}.
\label{eq:simpson_share}
\end{equation}
The index equals zero when all assets in a cluster come from the same class and reaches its maximum when the three asset classes are equally represented. For three equally represented classes, the theoretical maximum is \(D_{\max}=2/3\).

We summarise cross-class mixing by averaging over robust clusters larger than a minimum size threshold:
\begin{equation}
\langle D\rangle_t = \frac{1}{|\mathcal{C}^{\ast}_t|} \sum_{C\in\mathcal{C}^{\ast}_t} D_{C,t},
\label{eq:average_diversity}
\end{equation}
where \(\mathcal{C}^{\ast}_t\) denotes the set of robust clusters with more than five assets. A higher \(\langle D\rangle_t\) indicates weaker asset-class boundaries and greater compositional mixing inside robust communities.

\subsection{Asset data and market indicators}

We analyse the period from 15 October 2017 to 29 February 2024, yielding \(T_{\mathrm{total}}=2,329\) calendar days. The empirical setting combines three asset classes: cryptocurrencies, fiat currencies, and S\&P500 equities. This design allows us to compare the behaviour of a digital asset market, a global currency market, and a traditional equity market within a common cross-asset framework.

The raw cryptocurrency dataset contains daily prices recorded at midnight for 5,450 cryptocurrencies compiled from multiple public data sources (\url{www.investing.com}, \url{coinmarketcap.com}, \url{www.coindesk.com}, \url{www.coincodex.com}, and \url{www.marketwatch.com}). We first identified the top 1,000 cryptocurrencies by market capitalisation as of 22 February 2022 and retained those that were active over the full sample period. Cryptocurrencies with more than 10 missing daily prices were excluded, leaving 157 eligible cryptocurrencies before the final balancing step. The fiat dataset contains daily exchange rates against the U.S. dollar for more than 180 currencies, obtained from Yahoo Finance~\citep{aroussi2023yfinance}. After excluding currencies with more than 10 missing observations, 127 fiat currencies remained. The equity dataset contains daily closing prices for S\&P500 constituents over the same period, also obtained from Yahoo Finance~\citep{aroussi2023yfinance}. After excluding stocks with more than 10 missing observations, 484 stocks remained eligible. The 2022 ranking is used only to define the initial cryptocurrency universe, while the 2024 ranking is used for the final balanced sample selection.

To ensure balance across asset classes, we retained the same number of assets from each market. Specifically, we ranked eligible cryptocurrencies and S\&P500 stocks by market capitalisation on 29 February 2024 and retained the top 127 assets in each class. Together with the 127 fiat currencies passing the missing-data filter, this yields a balanced final sample of \(N_{\mathrm{crypto}}=127\), \(N_{\mathrm{fiat}}=127\), \(N_{\mathrm{S\&P500}}=127\), and therefore \(N=381\) assets in the combined cross-asset system. The balanced design prevents the system-wide turbulence measure and network summaries from being mechanically dominated by the asset class with the largest number of available securities. The full asset list is provided in the Supplementary Information. Because cryptocurrencies trade continuously, while fiat currencies and equities follow different trading calendars, all price series are aligned to a common daily calendar. For non-trading days in fiat currencies and S\&P500 equities, prices are linearly interpolated. This produces a common daily time index for the cross-asset return vectors and rolling network estimates. 

In addition to asset prices, we include three macro-financial indicators that capture broad market conditions. The CBOE Volatility Index (\url{cboe.com}) measures option-implied volatility in the U.S.\ equity market and is used as a proxy for risk sentiment. The broad U.S.\ dollar index (\url{fred.stlouisfed.org}) captures movements in the value of the U.S.\ dollar relative to other major currencies. The U.S. Economic Policy Uncertainty Index (\url{fred.stlouisfed.org}) captures policy-related uncertainty based on news coverage. All macro-financial indicators are obtained at daily frequency, and weekend values are filled by linear interpolation to match the common calendar used for the asset-price data.

\section{Results and Discussion}

\subsection{Evolution of cross-asset network structure}
\label{sec:CNS}

\begin{figure}[h!]
    \centering
    \includegraphics[width=0.95\textwidth]{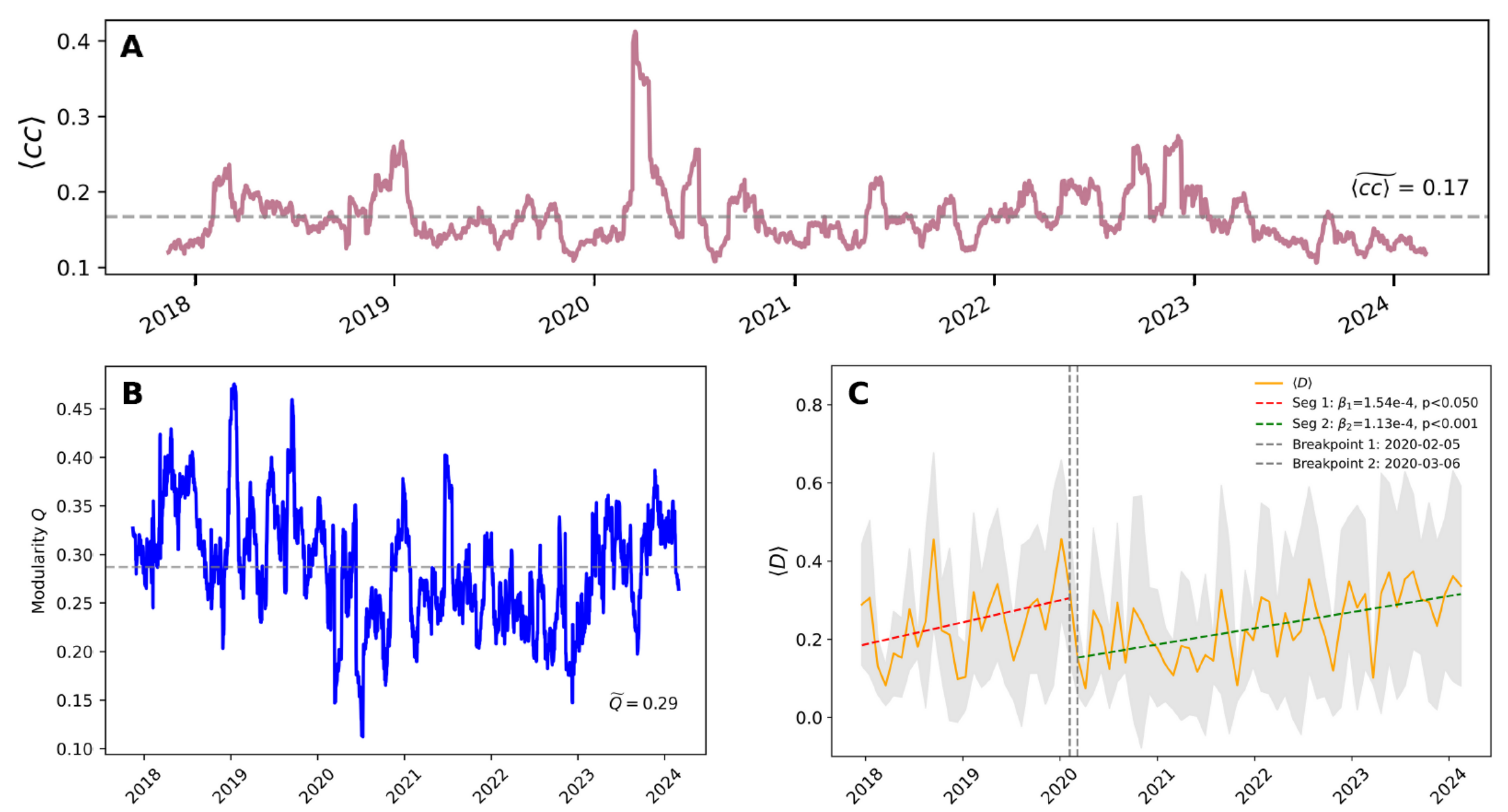}
    \caption{Evolution of network structure and modular organisation in the cross-asset network. Panel (A) shows the mean weighted clustering coefficient $\langle cc \rangle$, the dashed line indicates its average $\widetilde{\langle cc \rangle} = \frac{1}{T}\sum_{t=1}^{T} \langle cc \rangle_t$ over the entire period. Panel (B) shows the modularity $Q$ for the largest connected component and its average $\widetilde{Q}$. Panel (C) shows the average Simpson's Diversity Index $\langle D \rangle$ for stable large clusters with more than $5$ assets. The grey band indicates one standard deviation. The vertical dashed lines correspond to the estimated breakpoints of a piecewise linear regression, with slopes \(\beta\) for each segment.
    }
    \label{fig:network_evolution}
\end{figure}

Figure~\ref{fig:network_evolution}A shows that local cohesion varies episodically over time. The average weighted clustering coefficient fluctuates around its time average of \(\widetilde{\langle cc\rangle}=0.17\), increasing during periods of market stress. The largest spike occurs around the onset of the COVID-19 crisis in early 2020, indicating that assets became embedded in tighter local triangles of positive comovement. A second increase appears around late 2022, coinciding with the FTX-related disruption in the cryptocurrency market. These episodes suggest that stress temporarily compresses local market distances and strengthens short-run cross-asset cohesion, but the effect is not persistent, with clustering repeatedly returning toward, and later falling below, its long-run average. Figure~\ref{fig:network_evolution}B shows that modularity \(Q_t\) fluctuates around a time average of \(\widetilde{Q}=0.29\), alternating between periods of stronger compartmentalisation and stress-related integration. For example, the higher modularity in 2018--2019 indicates more compartmentalisation into internally cohesive communities. Around the COVID-19 period, modularity declines, indicating weaker community boundaries as cross-asset stress increased. It remains mostly below or near its long-run average from 2021 to 2022, before rising again in 2023 as the network becomes more segmented. Figure~\ref{fig:network_evolution}C shows that the average Simpson diversity index \(\langle D\rangle_t\) follows a pattern consistent with a gradual weakening of asset-class boundaries and remains below the theoretical maximum of \(D_{\max}=2/3\). Robust communities are not evenly balanced across cryptocurrencies, fiat currencies, and S\&P500 equities. Nevertheless, its growth over time indicates increasing cross-class mixing within stable communities. The piecewise linear fit suggests a stronger increase before the COVID-19 period, a temporary interruption around February--March 2020, and a slower positive trend afterwards. This pattern supports an episodic-integration interpretation, in which cross-asset markets remain partly segmented during calm periods, but stress episodes compress local distances, weaken modular separation, and increase cross-class mixing. Network structure is therefore an emergent descriptor of the market state, reflecting how asset-class boundaries and dependence structures reconfigure under stress rather than representing an exogenous source of turbulence.

\subsection{Market-specific turbulence and system-wide regimes}
\label{sec:MTR}

\begin{figure}[h!]
    \centering
    \includegraphics[width=1\textwidth]{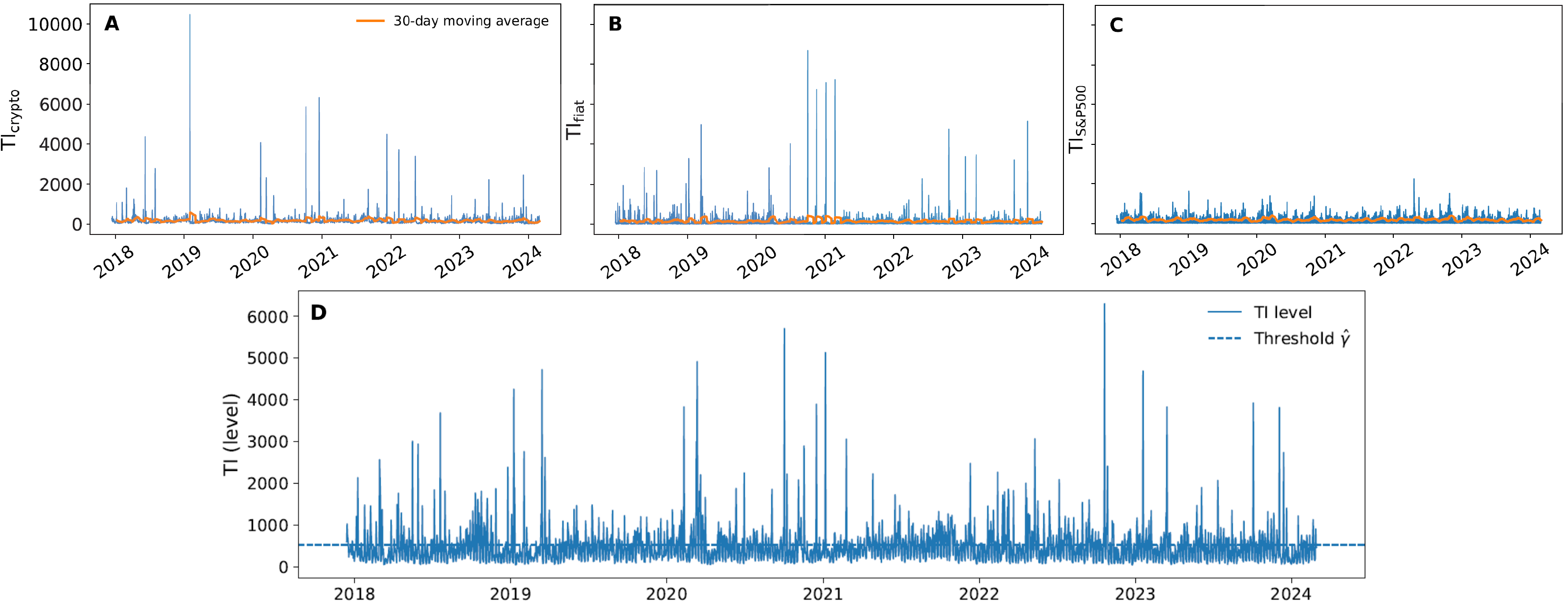}
    \caption{Evolution of market-specific and system-wide turbulence. All Turbulence Indices are computed with \(\Delta_{\mathrm{TI}}=30\). Turbulence Index for (A) the cryptocurrency, (B) fiat, and (C) S\&P500 markets, together with the 30-day moving average. (D) system-wide Turbulence Index computed from the full cross-asset return vector. The dashed line in panel (D) indicates the estimated threshold \(\hat{\gamma}\). Observations are assigned to the high-turbulence regime when \(\mathrm{TI}_{t-1}>\hat{\gamma}\), or to the low-turbulence regime otherwise (see Sec.~\ref{sec:methods_dynamic_transmission}).}
    \label{fig:Turbulence_Index}
\end{figure}

Figure~\ref{fig:Turbulence_Index}A--C show that the three market-specific indices exhibit distinct turbulence profiles. Cryptocurrency turbulence is highly episodic, with long intervals of moderate values interrupted by sharp peaks around 2019 and again during 2021--2022. Since the Turbulence Index measures covariance-adjusted displacement from recent market behaviour, these peaks indicate periods when cryptocurrency returns moved unusually far from their recent multivariate pattern. This suggests that digital-asset stress is concentrated in discrete bursts rather than evolving smoothly over time. Fiat turbulence also exhibits pronounced peaks, especially around the 2020--2021 period, when global currency markets were affected by pandemic-related uncertainty, uneven recovery dynamics, and shifts in monetary policy expectations. By contrast, S\&P500 turbulence is less dominated by isolated extreme peaks and instead displays shorter and more frequent fluctuations.

Figure~\ref{fig:Turbulence_Index}D shows the system-wide Turbulence Index used to define the low- and high-turbulence regimes. This index is computed from the full cross-asset return vector rather than by aggregating the three market-specific indices, so it rises when the combined system deviates from its recent multivariate benchmark. The estimated threshold, obtained from the TVAR specification (Section~\ref{sec:methods_dynamic_transmission}), lies above typical day-to-day variation but is crossed repeatedly during periods of market stress. The high-turbulence regime thus captures recurrent episodes of elevated system-wide displacement, rather than only a small number of extreme outliers or single-market volatility events. This regime definition is used in the state-dependent transmission analysis in the next section.

\subsection{Distributional properties and stationarity diagnostics}

\begin{table}[htbp]
\centering
\setlength{\tabcolsep}{10pt}  
\begin{tabular}{lrrrrrr}
\toprule
\textbf{Variable} & \textbf{Mean} & \textbf{Std. Dev.} & \textbf{Min} & \textbf{Max} & \textbf{Skewness} & \textbf{Excess Kurtosis} \\
\midrule
$\boldsymbol{\text{TI}_{\textbf{s\&p500}}}$ & 185.54 & 219.07 & 3.71 & 2259.59 & 2.15 & 8.33 \\
$\boldsymbol{\text{TI}_{\textbf{fiat}}}$ & 157.64 & 440.79 & 3.24 & 8689.96 & 11.46 & 165.96 \\
$\boldsymbol{\text{TI}_{\textbf{crypto}}}$ & 184.99 & 373.78 & 10.95 & 10481.77 & 15.73 & 334.81 \\
\textbf{$\left\langle \boldsymbol{cc} \right\rangle$} & 0.17 & 0.04 & 0.11 & 0.41 & 2.10& 7.67 \\
\textbf{Q} & 0.29 & 0.06 & 0.11 & 0.48 & 0.26 & 0.09 \\
\textbf{$\left\langle \boldsymbol{D} \right\rangle$} & 0.26 & 0.08 & 0.07 & 0.54 & 0.45 & 0.20 \\ 
\textbf{VIX} & 20.27 & 7.91 & 9.15 & 82.69 & 2.44 & 10.92 \\
\textbf{USDX} & 116.70 & 4.33 & 106.49 & 128.45 & 0.17 & -0.25 \\
\textbf{EPU} & 151.48 & 111.18 & 4.05 & 1026.38 & 2.39 & 7.83 \\
\bottomrule
\end{tabular}
\caption{Descriptive statistics of the daily variables in levels. Skewness and excess kurtosis are computed from sample moments.}
\label{tab:desc_stats}
\end{table}

The three market turbulence indices are strongly right-skewed and leptokurtic, indicating that turbulence is concentrated in irregular bursts rather than distributed evenly over time (Table~\ref{tab:desc_stats}). This pattern is more prominent for cryptocurrency turbulence, which has the largest maximum, skewness, and excess kurtosis among the market-specific indices. Fiat turbulence also shows substantial tail risk, while S\&P500 turbulence is less extreme by comparison, suggesting a more compressed distribution of multivariate equity-market stress. The network indicators, average clustering coefficient \(\langle cc\rangle_t\), modularity \(Q_t\), and diversity index \(\langle D\rangle_t\), are bounded by construction and vary within narrower ranges than the turbulence indices, indicating smoother changes in network structure. Return-based stress can produce sharp short-run displacements, whereas changes in local cohesion, modular segmentation, and cross-class mixing are more gradual. Among the macro-financial variables, VIX and EPU are positively skewed and heavy-tailed, reflecting occasional surges in volatility expectations and policy uncertainty, whereas USDX is more symmetric and less spike-driven over the full sample.

\begin{table}[htbp]
\centering
\setlength{\tabcolsep}{10pt}       
\begin{tabular}{lcccc}
\toprule
 & \multicolumn{2}{c}{\textbf{Levels}} & \multicolumn{2}{c}{\textbf{First Differences}} \\
\cmidrule(lr){2-3} \cmidrule(lr){4-5}
\textbf{Variable} & \textbf{ADF} & \textbf{KPSS} & \textbf{ADF} & \textbf{KPSS} \\
\midrule
$\boldsymbol{\text{TI}_{\textbf{s\&p500}}}$ & \textless 0.001 & $>0.10$ & \textless 0.001 & $>0.10$ \\
$\boldsymbol{\text{TI}_{\textbf{fiat}}}$      & \textless 0.001 & $>0.10$ & \textless 0.001 & $>0.10$ \\
$\boldsymbol{\text{TI}_{\textbf{crypto}}}$    & \textless 0.001 & $>0.10$ & \textless 0.001 & $>0.10$ \\
\textbf{$\left\langle \boldsymbol{cc} \right\rangle$} & \textless 0.001 & $>0.10$ & \textless 0.001 & $>0.10$ \\
\textbf{Q}                                    & \textless 0.001 & $<0.01$ & \textless 0.001 & $>0.10$ \\
\textbf{$\left\langle \boldsymbol{D} \right\rangle$}  & \textless 0.001 & $>0.10$ & \textless 0.001 & $>0.10$ \\  
\textbf{VIX}                                  & \textless 0.001 & 0.024   & \textless 0.001 & $>0.10$ \\
\textbf{USDX}                                 & 0.257              & $<0.01$ & \textless 0.001 & $>0.10$ \\
\textbf{EPU}                                  & 0.074             & 0.017   & \textless 0.001 & $>0.10$ \\
\bottomrule
\end{tabular}
\caption{Unit root diagnostics for the variables entering the VAR system. The table reports p-values from Augmented Dickey--Fuller (ADF) and Kwiatkowski--Phillips--Schmidt--Shin (KPSS) tests in levels and first differences.}
\label{tab:unitroot}
\end{table}

The ADF test rejects the unit-root null for most variables in levels, but not for USDX, and only weakly for EPU (Table~\ref{tab:unitroot}). The KPSS test rejects level stationarity for \(Q_t\), VIX, USDX, and EPU. Estimating the VAR directly in levels would thus combine variables with different persistence properties. For first differences, ADF rejects the unit-root null, while KPSS fails to reject stationarity for all variables. The dynamic transmission model is therefore estimated using standardised first differences (Eq.~\eqref{eq:standardised_difference}). This specification focuses the VAR on changes in market turbulence, network structure, and macro-financial conditions, while placing all variables on a comparable scale.

\subsection{Full-sample connectedness and impulse responses}

\begin{table}
\centering
\begin{tabular}{lrrrrrrrrr|r}
\toprule
 & $\boldsymbol{\text{TI}_{\textbf{s\&p500}}}$ & $\boldsymbol{\text{TI}_{\textbf{fiat}}}$ & $\boldsymbol{\text{TI}_{\textbf{crypto}}}$ & \textbf{$\left\langle \boldsymbol{cc} \right\rangle$} & \textbf{Q} & \textbf{$\left\langle \boldsymbol{D} \right\rangle$} & \textbf{VIX} & \textbf{USDX} & \textbf{EPU} & \textbf{From} \\
\midrule
$\boldsymbol{\text{TI}_{\textbf{s\&p500}}}$  & 92.505 & 3.263 & 0.393 & 0.317 & 0.345 & 0.372 & 1.067 & 0.382 & 1.356 & 7.495 \\
$\boldsymbol{\text{TI}_{\textbf{fiat}}}$ & 2.872 & 94.472 & 0.038 & 0.063 & 0.167 & 0.357 & 0.696 & 0.754 & 0.581 & 5.528 \\
$\boldsymbol{\text{TI}_{\textbf{crypto}}}$ & 0.600 & 1.798 & 95.756 & 0.531 & 0.386 & 0.152 & 0.161 & 0.504 & 0.112 & 4.244 \\
\textbf{$\left\langle \boldsymbol{cc} \right\rangle$}  & 0.448 & 0.646 & 0.282 & 75.593 & 14.543 & 1.196 & 6.491 & 0.517 & 0.284 & 24.407 \\
\textbf{Q} & 0.756 & 0.247 & 0.371 & 15.563 & 81.631 & 0.246 & 0.523 & 0.240 & 0.423 & 18.369 \\
\textbf{$\left\langle \boldsymbol{D} \right\rangle$} & 0.271 & 0.295 & 1.060 & 1.035 & 0.689 & 95.219 & 0.875 & 0.158 & 0.397 & 4.780 \\
\textbf{VIX} & 1.937 & 0.553 & 0.253 & 9.551 & 1.556 & 0.736 & 79.527 & 5.506 & 0.380 & 20.472 \\
\textbf{USDX} & 0.306 & 0.540 & 0.126 & 1.424 & 0.529 & 0.209 & 8.606 & 87.775 & 0.485 & 12.225 \\
\textbf{EPU} & 2.154 & 1.374 & 0.302 & 0.298 & 0.248 & 0.201 & 0.689 & 0.434 & 94.300 & 5.700 \\
\midrule
\textbf{To} & 9.344 & 8.716 & 2.825 & 28.782 & 18.463 & 3.469 & 19.108 & 8.495 & 4.018 &  \\
\textbf{Net} & 1.849 & 3.188 & -1.419 & 4.375 & 0.094 & -1.311 & -1.364 & -3.730 & -1.682 & \textbf{11.469} \\
\bottomrule
\end{tabular}
\caption{Full-sample Diebold--Yilmaz connectedness table based on the generalised forecast error variance decomposition (GFEVD) at horizon \(H=30\). Entries report the percentage of the 30-day-ahead forecast error variance of the row variable attributable to shocks in the column variable. \textbf{From}, \textbf{To}, and \textbf{Net} denote received, transmitted, and net spillovers, respectively. The bottom-right entry reports the total spillover index. All entries are percentages.}
\label{tab:dy_full}
\end{table}

Table~\ref{tab:dy_full} reports the Diebold--Yilmaz connectedness decomposition at a 30-day horizon. Each row decomposes the forecast uncertainty of a given variable into contributions from its own shock and from shocks originating elsewhere in the system. Since columns represent shock origins and rows represent affected variables, the table is directional and not symmetric.

The diagonal entries dominate, indicating that most forecast uncertainty is self-driven in the full-sample VAR. This is especially clear for the three market turbulence indices, where own shocks explain \(92.51\%\) of S\&P500 turbulence, \(94.47\%\) of fiat turbulence, and \(95.76\%\) of cryptocurrency turbulence. Nevertheless, the off-diagonal entries reveal selective cross-market transmission. S\&P500 turbulence contributes more to fiat and cryptocurrency turbulence than it receives from them, while the reverse contribution from cryptocurrency turbulence to traditional markets is smaller. Consistent with this asymmetry, S\&P500 and fiat turbulence are net transmitters, with net positions of \(1.85\) and \(3.19\) percentage points, whereas cryptocurrency turbulence is a net receiver with \(-1.42\) percentage points. This indicates that cryptocurrency turbulence is integrated into the broader system but is not the dominant transmitter in the average full-sample dynamics. 

The average clustering coefficient \(\langle cc\rangle\) receives \(24.41\%\) of its forecast uncertainty from the rest of the system and transmits \(28.78\%\), giving the largest positive net position in the table. This suggests that local network cohesion is both shaped by system-wide conditions and linked to subsequent forecast uncertainty elsewhere in the system. Modularity \(Q\) is closer to balance, while the diversity index \(\langle D\rangle\) is relatively self-driven, with \(95.22\%\) of its variance explained by own shocks. Thus, local clustering is more involved in short-run transmission than global compositional mixing.

Among macro-financial variables, VIX is the most connected. It receives \(20.47\%\) of its forecast uncertainty from the system and transmits \(19.11\%\), giving a near-balanced but slightly negative net position. VIX contributes \(6.49\%\) to the forecast uncertainty of \(\langle cc\rangle\), whereas \(\langle cc\rangle\) contributes \(9.55\%\) to VIX. The first channel is consistent with implied volatility being followed by tighter local comovement in the cross-asset network. The reverse channel suggests that network clustering contains predictive information about subsequent variation in implied volatility, possibly because broader return synchronisation is related to the covariance environment relevant for portfolio risk and option pricing. The GFEVD cannot identify a structural mechanism, but it indicates a non-trivial predictive link between local clustering and volatility uncertainty.

The total spillover index is \(11.47\%\), meaning that about one-ninth of the system's 30-day forecast uncertainty is explained by cross-variable shocks rather than own shocks. The full-sample system is therefore not dominated by pervasive cross-variable spillovers. Instead, connectedness is moderate and selective, with economically meaningful transmission concentrated around S\&P500 and fiat turbulence, VIX, and the network clustering coefficient.

\begin{figure}[h!]
    \centering
    \includegraphics[width=1\textwidth]{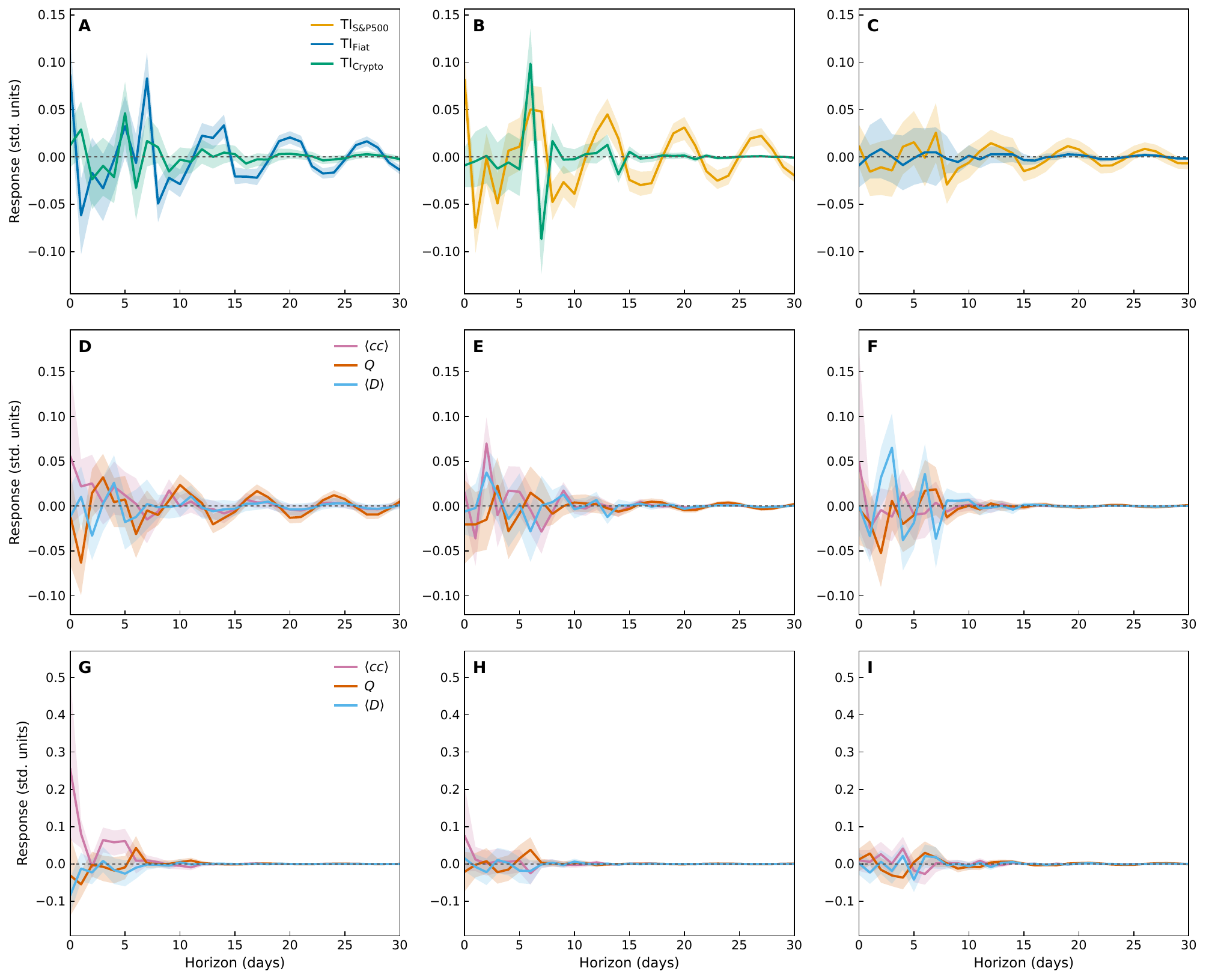}  
    \caption{Generalised impulse responses across the market--network--macro system. Rows correspond to response groups and columns to originating shocks. The first row reports cross-market responses among the three market turbulence indices following shocks to (A) S\&P500 turbulence, (B) fiat turbulence, and (C) cryptocurrency turbulence, with own responses suppressed. The second row reports responses of network indicators to shocks in (D) S\&P500 turbulence, (E) fiat turbulence, and (F) cryptocurrency turbulence. The third row reports network responses to shocks in (G) VIX, (H) USDX, and (I) EPU. Responses are computed using the Pesaran--Shin generalised framework, expressed in standard-deviation units. Shaded bands denote 90\% bootstrap C.I.s based on \(B=300\) residual-bootstrap replications.}
    \label{fig:girfs_networks}
\end{figure}

The connectedness table reports forecast-uncertainty shares, while the GIRFs trace the response path after a one-standard-deviation unexpected change over subsequent days. The temporal responses are estimated from the full-sample VAR and should therefore be interpreted as average dynamic responses over the sample, not as responses during a specific crisis (Fig.~\ref{fig:girfs_networks}). A one-standard-deviation shock represents an unexpected increase in the standardised first difference of the originating variable. A shock to S\&P500 turbulence generates a short-run response in fiat turbulence and a weaker response in cryptocurrency turbulence (Fig.~\ref{fig:girfs_networks}A). A shock to fiat turbulence also affects S\&P500 turbulence and produces a delayed response in cryptocurrency turbulence (Fig.~\ref{fig:girfs_networks}B). By contrast, shocks originating in cryptocurrency turbulence in Fig.~\ref{fig:girfs_networks}C have limited effects on the two traditional markets. This pattern is consistent with the connectedness table, i.e.\ cryptocurrency turbulence responds to the broader system, but it is not the dominant source of full-sample cross-market transmission.

Market stress can disturb several layers of network organisation, but the affected layer and the duration of the response differ by the originating market (Fig.~\ref{fig:girfs_networks}D--F). Following an S\&P500 turbulence shock in Fig.~\ref{fig:girfs_networks}D, the network response is relatively long-lasting and continues to fluctuate before dissipating. This suggests that equity-market turbulence produces a more prolonged disturbance to community organisation and local cohesion, even though it does not lead to a permanent topological shift. A fiat turbulence shock in Fig.~\ref{fig:girfs_networks}E produces a sharper but shorter-lived adjustment across \(\langle cc\rangle\) and \(\langle D\rangle\), indicating that currency-market stress temporarily affects local cohesion and cross-class composition, but is absorbed relatively quickly. A cryptocurrency turbulence shock generates an early response in \(\langle D\rangle\), indicating a temporary disturbance in cross-class community composition rather than a sustained change in local clustering or modular separation (Fig.~\ref{fig:girfs_networks}F). Market-to-network transmission is thus distributed across different dimensions of the dependence network.

A VIX shock produces the clearest response, with stronger local clustering and weaker modular separation at short horizons (Fig.~\ref{fig:girfs_networks}G). This pattern is consistent with a flight-to-correlation mechanism, in which higher volatility expectations are followed by a more compressed dependence structure and weaker separation between previously distinct communities. Responses to USDX and EPU shocks are weaker and less persistent (Fig.~\ref{fig:girfs_networks}H--I), suggesting that broad dollar movements and policy uncertainty affect the network less directly than volatility expectations in the full-sample VAR. Network structure acts as a short-run reconfiguration channel in the average linear system. Market and macro-financial shocks temporarily disturb local cohesion, modular organisation, and cross-class composition, but the responses revert toward zero.

\subsection{Rolling connectedness and net transmission}

The full-sample connectedness estimates describe average spillovers over the entire sample. To examine whether this average masks changes in aggregate transmission and transmitter--receiver roles, we estimate rolling-window connectedness measures. This distinguishes moderately connected periods from episodes in which forecast uncertainty is transmitted more widely across market turbulence, network structure, and macro-financial variables.

\begin{figure}[h!]
    \centering
    \includegraphics[width=0.8\textwidth]{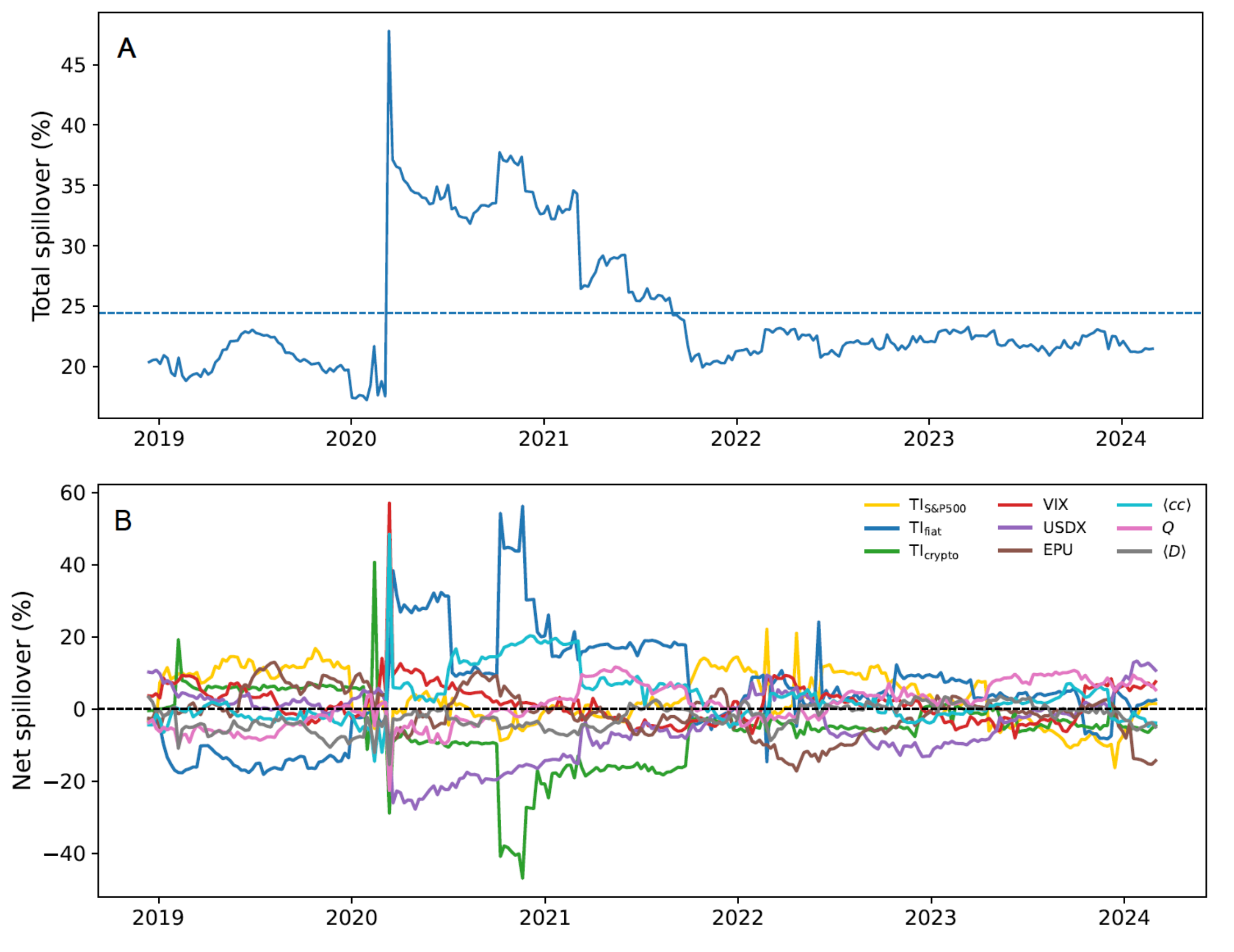} 
     \caption{Rolling aggregate and directional net spillover dynamics based on the Diebold--Yilmaz generalised FEVD framework. (A) Rolling Total Spillover Index (TSI), estimated from a VAR fitted on a 365-day rolling window updated weekly with forecast horizon \(H=30\). The dashed horizontal line denotes the sample mean. (B) reports rolling net connectedness, defined as transmitted minus received spillovers.}
    \label{fig:Rolling_TSI}
\end{figure}

Figure~\ref{fig:Rolling_TSI}A shows substantial time variation in aggregate connectedness. The rolling TSI ranges from \(17.21\%\) to \(47.80\%\), with a sample mean of \(24.42\%\), which is higher than the full-sample TSI reported in Table~\ref{tab:dy_full}. This difference indicates that the full-sample VAR smooths over periods in which cross-variable transmission is much stronger. Before the COVID-19 crisis, the TSI remains in a relatively narrow band around \(20\%\), suggesting a moderate baseline level of cross-variable dependence. At the onset of the pandemic, the index rises sharply and reaches its maximum of \(47.80\%\) on 12 March 2020, indicating that forecast uncertainty became much more shared across market turbulence, network structure, and macro-financial variables. The index remains elevated through much of 2020 and early 2021 before declining toward a lower range from late 2021 onward. Thus, aggregate connectedness is episodic, i.e.\ it intensifies sharply during broad stress events but does not remain permanently high.

By decomposing aggregate connectedness into time-varying net spillover positions, we see that transmission roles are unstable over time, rather than fixed properties of individual variables (Fig.~\ref{fig:Rolling_TSI}B). Fiat and cryptocurrency turbulence display the largest swings. Fiat turbulence moves from a net receiver before the pandemic to a major transmitter during the high-connectedness period of 2020--2021, whereas cryptocurrency turbulence becomes strongly negative during the same period. This suggests that currency-market turbulence became a central propagation channel during the pandemic-related adjustment, while cryptocurrency turbulence mainly absorbed forecast uncertainty from the broader system. The macro-financial variables show a different pattern. VIX displays a sharp transmitter episode around the onset of COVID-19, consistent with volatility expectations becoming more central to short-run transmission. USDX and EPU are more often negative, indicating that dollar movements and policy uncertainty tend to absorb more forecast uncertainty than they transmit in the rolling system. The network measures indicate that \(\langle cc\rangle\) becomes positive around the onset of COVID-19 and again later in the sample, linking stronger local network cohesion to outward transmission during specific stress episodes. By contrast, \(Q\) and \(\langle D\rangle\) fluctuate closer to zero for much of the sample, so global community separation and cross-class mixing are less frequently dominant rolling transmitters. Connectedness is thus both time-varying and selective. Stress episodes increase aggregate spillovers, but they also change the identity of transmitters and receivers. This calendar-time instability motivates the regime analysis below, which separates observations by turbulence state rather than by rolling historical windows.

\subsection{Regime-dependent connectedness and network reconfiguration}

To separate calendar-time variation from state dependence, we estimate a two-regime (calm and stressed market) threshold VAR defined by the system-wide Turbulence Index. Observations are assigned to the low-turbulence regime if \(\mathrm{TI}_{t-1}\leq \hat{\gamma}\), and to the high-turbulence regime if \(\mathrm{TI}_{t-1}>\hat{\gamma}\). This allows the VAR coefficients, the residual covariance matrix, the connectedness measures, and the impulse responses to vary across regimes. The purpose is to assess whether stress changes both spillover magnitudes and the organisation of transmitter--receiver roles. Stability diagnostics reported in the Supplementary Information confirm that both regime-specific VARs admit well-defined moving-average representations.

\begin{table}
\centering
\begin{tabular}{l|rrr|rrr|r}
\toprule
 & \textbf{To (Low)} & \textbf{From (Low)} & \textbf{Net (Low)} & \textbf{To (High)} & \textbf{From (High)} & \textbf{Net (High)} & \textbf{$\Delta$Net (High-Low)}  \\
\midrule
$\boldsymbol{\text{TI}_{\textbf{s\&p500}}}$ & 14.645 & 8.096 & 6.549 & 14.864 & 26.565 & -11.700 & -18.249 \\
$\boldsymbol{\text{TI}_{\textbf{fiat}}}$  & 8.421 & 12.186 & -3.765 & 54.743 & 11.191 & 43.552 & 47.317 \\
$\boldsymbol{\text{TI}_{\textbf{crypto}}}$ & 1.708 & 2.250 & -0.542 & 16.209 & 12.170 & 4.040 & 4.582 \\
\textbf{$\left\langle \boldsymbol{cc} \right\rangle$} & 11.915 & 13.793 & -1.878 & 47.640 & 39.959 & 7.681 & 9.559 \\
\textbf{Q} & 11.032 & 11.879 & -0.846 & 30.954 & 28.944 & 2.010 & 2.856 \\
\textbf{$\left\langle \boldsymbol{D} \right\rangle$} & 3.672 & 3.724 & -0.052 & 7.021 & 16.405 & -9.384 & -9.332 \\
\textbf{VIX} & 15.459 & 7.133 & 8.326 & 31.571 & 40.676 & -9.105 & -17.431 \\
\textbf{USDX} & 3.460 & 10.259 & -6.799 & 13.432 & 24.216 & -10.785 & -3.986 \\
\textbf{EPU} & 7.244 & 8.237 & -0.993 & 2.675 & 18.985 & -16.309 & -15.316 \\
\bottomrule
\end{tabular}
\caption{Regime-specific directional connectedness under the threshold VAR. \textbf{To} and \textbf{From} denote transmitted and received spillovers, respectively. \textbf{Net} is \textbf{To}$-$\textbf{From}, and \(\Delta\)\textbf{Net} is the change in net connectedness from the low- to the high-turbulence regime. \textbf{To} and \textbf{From} are reported as percentages of forecast error variance, and \textbf{Net} and \(\Delta\)\textbf{Net} are reported in percentage points.}
\label{tab:tvar_tofromnet}
\end{table}

Table~\ref{tab:tvar_tofromnet} shows that transmitter--receiver roles change substantially across regimes. In the low-turbulence regime, net transmission is concentrated in S\&P500 turbulence and VIX, with net positions of \(6.55\) and \(8.33\) percentage points, respectively. Most other variables are weak or neutral receivers, and the network indicators are slightly negative or near zero. In these periods, network structure mainly reflects market organisation rather than acting as an outward transmitter of forecast uncertainty. The high-turbulence regime, on the other hand, shows a different transmission structure. Fiat turbulence becomes the dominant net transmitter, with \(\mathrm{Net}=43.55\) percentage points and \(\Delta\mathrm{Net}=47.32\) percentage points, while S\&P500 turbulence switches from a net transmitter to a net receiver. Cryptocurrency turbulence becomes mildly transmissive, but its role remains much smaller compared to fiat turbulence. This shift is consistent with the idea that currency-market turbulence captures stress in global funding conditions, monetary-policy expectations, and cross-border risk repricing. During elevated system-wide turbulence, exchange-rate movements may therefore transmit macro-financial pressure across equity and cryptocurrency markets. The VAR cannot identify the underlying structural mechanism, but the regime contrast shows that fiat currency-market turbulence is not simply a control variable but becomes the central stress-state transmitter in the cross-asset system.

Network structure also becomes more active in the high-turbulence regime. The clustering coefficient \(\langle cc\rangle\) moves from a mild net receiver to a net transmitter, with \(\Delta\mathrm{Net}=9.56\) percentage points, while modularity \(Q\) also becomes mildly positive. This indicates that tighter local comovement and changes in community organisation are not simply a result of stress. Under elevated turbulence, they become more involved in transmitting forecast uncertainty. On the other hand, diversity \(\langle D\rangle\) becomes more negative, implying that cross-class compositional mixing is mainly shaped by shocks from the broader system rather than acting as an independent transmitter. The high-turbulence regime therefore combines a market channel, dominated by fiat currency turbulence, with a structural channel centred on local clustering and modular organisation.

The macro-financial variables become net receivers in the high-turbulence regime. VIX, USDX, and EPU all have negative net positions, with EPU showing the strongest receiver role. This differs from the rolling analysis, where VIX appears as a transmitter around the COVID-19 period, because the two models condition on different dimensions of the data. The rolling VAR tracks calendar-time variation, whereas the threshold VAR separates observations by turbulence state.

\begin{table}[tbp]
\centering
\begin{threeparttable}
\begin{tabular}{
    ll
    S[table-format=2.3]
    S[table-format=2.3]
    S[table-format=2.3]
}
\toprule
\multicolumn{5}{c}{\textbf{Block spillover matrices}} \\
\midrule
\textbf{Regime} & \textbf{Block}  & {\textbf{Markets}} & {\textbf{Network}} & {\textbf{Macro}} \\
\midrule
\multirow{3}{*}{Low}
    & Markets & 96.540 & 0.821 & 2.639 \\
    & Network & 1.406 & 96.939 & 1.654 \\
    & Macro   & 2.801 & 1.314 & 95.884 \\
\addlinespace[2pt]
\multirow{3}{*}{High}
    & Markets & 93.326 & 3.921 & 2.753 \\
    & Network & 7.513 & 86.461 & 6.026 \\
    & Macro   & 11.125 & 9.720 & 79.154 \\
\midrule
\multicolumn{5}{c}{\textbf{Directional block connectedness}} \\
\midrule
\textbf{Regime} & \textbf{Block} & {\textbf{From}} & {\textbf{To}} & {\textbf{Net}} \\
\midrule
\multirow{3}{*}{Low}
    & Markets & 3.460 & 4.207 & 0.747 \\
    & Network & 3.061 & 2.135 & -0.926 \\
    & Macro   & 4.116 & 4.294 & 0.178 \\
\addlinespace[2pt]
\multirow{3}{*}{High}
    & Markets & 6.674 & 18.638 & 11.964 \\
    & Network & 13.539 & 13.642 & 0.103 \\
    & Macro   & 20.846 & 8.779 & -12.066 \\
\bottomrule
\end{tabular}
\end{threeparttable}
\caption{Regime-dependent block spillovers under the threshold VAR. The upper panel reports block-level GFEVD shares for market turbulence, network structure, and macro-financial conditions. Rows denote receiving blocks and columns transmitting blocks. The lower panel reports directional block connectedness, where \textbf{From} is the total received spillover, \textbf{To} is the total transmitted spillover, and \(\textbf{Net}=\textbf{To}-\textbf{From}\). Matrix entries, \textbf{From}, and \textbf{To} are percentages of forecast error variance, with \textbf{Net} reported in percentage points.}
\label{tab:tvar_block_combined}
\end{table}

Table~\ref{tab:tvar_block_combined} aggregates the regime-specific connectedness measures into three blocks: Markets, Network, and Macro. In the low-turbulence regime, the block spillover matrix is close to diagonal. Each block explains more than \(95\%\) of its own forecast error variance, and cross-block spillovers are small. The directional block measures are also close to zero, indicating that no block strongly dominates transmission under calm conditions. In the high-turbulence regime, the Market block becomes the dominant net transmitter, with \(\mathrm{Net}=11.96\) percentage points, whereas the Macro block becomes the dominant net receiver, with \(\mathrm{Net}=-12.07\) percentage points. The Network block remains close to balanced in net terms, but both its received and transmitted spillovers increase. This distinction indicates that the network block is not the main net transmitter at the aggregate level, but it becomes more tightly coupled to the rest of the system under stress.

The largest changes occur in the Market--Network and Network--Macro links. The share of Network forecast uncertainty explained by Markets rises from \(1.41\%\) to \(7.51\%\), indicating that market turbulence reshapes network structure more strongly in the high regime. At the same time, the share of Macro forecast uncertainty explained by Network rises from \(1.31\%\) to \(9.72\%\), linking network reconfiguration more strongly to the broader stress-transmission environment. High turbulence is therefore not simply a period of larger within-market spillovers, but also a regime in which markets, network structure, and macro-financial conditions become more tightly coupled.

\begin{figure}[h!]
    \centering
    \includegraphics[width=1\textwidth]{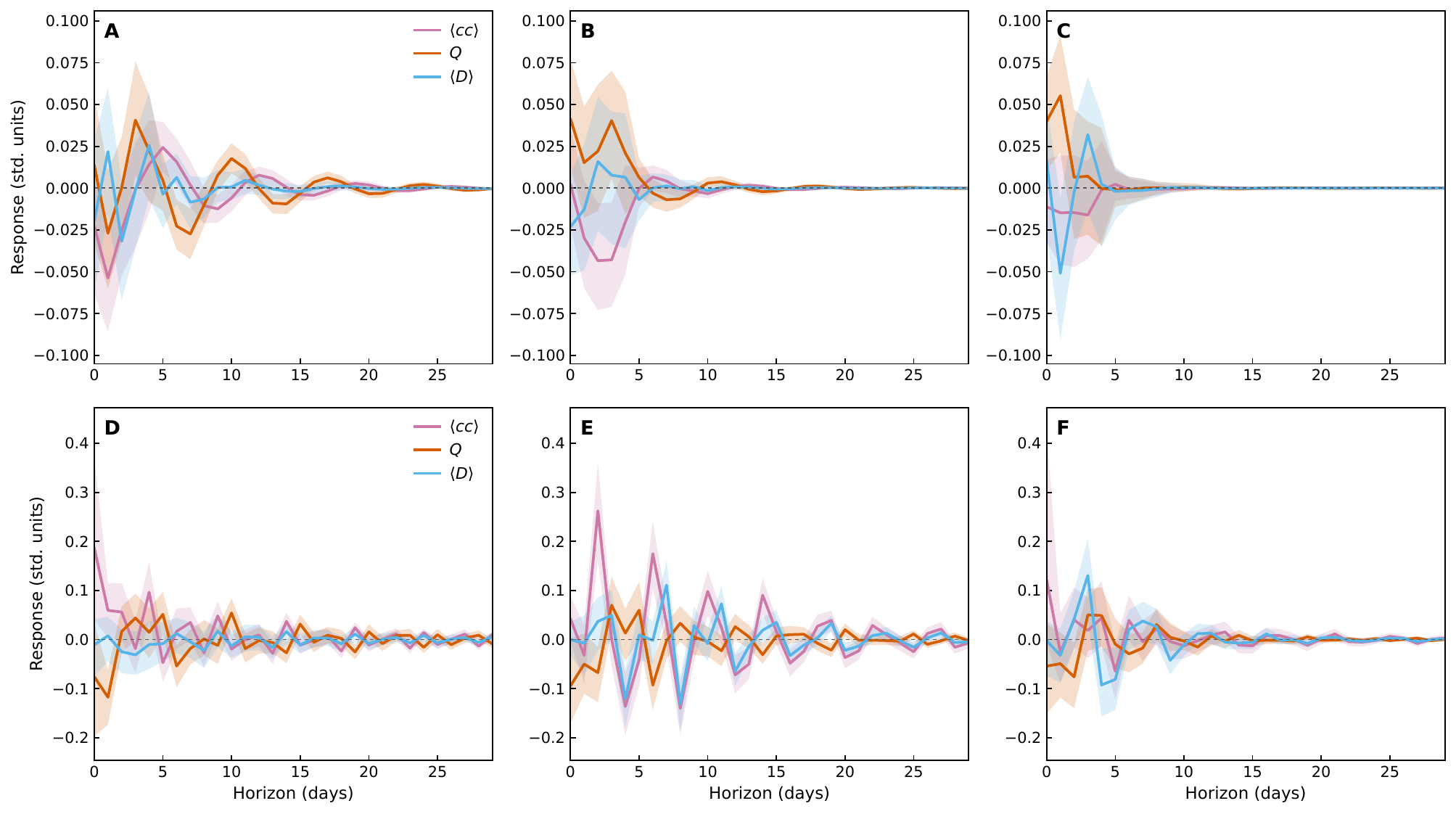}  
    \caption{Regime-conditional generalised impulse response functions from market turbulence shocks to network structure. Rows correspond to the low- and high-turbulence regimes identified by the threshold VAR, and columns correspond to the originating market-turbulence shock. Responses of (A) \(\langle cc\rangle\), (B) \(Q\), and (C) \(\langle D\rangle\) to shocks in S\&P500, fiat, and cryptocurrency turbulence under the low-turbulence regime. Panels (D)--(F) show the corresponding responses under the high-turbulence regimes. Responses are expressed in standardised units and traced over \(H=30\) days. Shaded areas denote 90\% bootstrap C.I.s from a regime-conditional residual bootstrap with \(B=300\) replications. }
    \label{fig:GIRF_LowHigh}
\end{figure}

We use regime-conditional GIRFs to examine how market shocks translate into changes in network structure within each regime.
In the low-turbulence regime, responses of \(\langle cc\rangle\), \(Q\), and \(\langle D\rangle\) remain small, indicating that ordinary market shocks do not substantially reorganise the cross-asset dependence network (Fig.~\ref{fig:GIRF_LowHigh}(A)--(C)). The network absorbs these shocks with limited movement in local cohesion, modular separation, or cross-class composition. In the high-turbulence regime, the same type of market shock produces much larger responses (Fig.~\ref{fig:GIRF_LowHigh}(D)--(F)). The amplification is most visible for \(\langle cc\rangle\), especially after S\&P500 and fiat turbulence shocks, indicating that local triangles of comovement tighten when the system is already stressed. This indicates that assets already close in the dependence network move more synchronously following an additional market turbulence shock. The response of modularity \(Q\) captures the mesoscopic organisation of the network. Short-run movements in \(Q\) under the high regime indicate that market shocks disturb community separation, rather than only strengthening pairwise correlations. Declines in \(Q\) correspond to weaker separation between previously distinct communities, while rebounds indicate partial recovery of segmented organisation. Elevated turbulence, therefore, affects not only the intensity of links but also the organisation of asset groups.

The weaker response of \(\langle D\rangle\) indicates that high-turbulence shocks affect network structure unevenly. Local clustering and modularity respond more strongly because they capture the intensity and organisation of links inside the dependence network. Cross-class diversity, by contrast, reflects the composition of robust communities and adjusts more slowly. Market shocks therefore first disturb the strength and mesoscopic organisation of connections, while changes in the asset-class composition of communities are less immediate. This regime contrast helps explain why network variables become more involved in the high-turbulence connectedness results. In calm states, the network structure primarily reflects the prevailing organisation of cross-asset dependence. In high-turbulence states, market shocks generate larger short-run movements in local clustering and modular structure, making network structure more involved in spillover propagation. The responses remain temporary, but their larger magnitude under stress supports a state-dependent interpretation of network structure rather than a persistent exogenous source of turbulence.

\section{Conclusion}

This study shows that cross-asset risk is not captured only by higher volatility or stronger average correlations. It is also reflected in how the dependence structure linking cryptocurrencies, fiat currencies, and S\&P500 equities reorganises under stress. Using rolling correlation networks, consensus-based community detection, market-specific and system-wide Turbulence Indices, and VAR-based connectedness analysis, we document an episodic form of cross-asset integration. In calm periods, the combined system remains partly segmented, with a modular structure and robust communities only partially mixing asset classes. Most forecast uncertainty is explained by own-market dynamics in calm periods. Under elevated turbulence, however, local clustering increases, modular separation weakens, and communities become more compositionally mixed, indicating compressed asset-class boundaries.

The dynamic transmission results show that this structural reorganisation is also linked to the transmission of forecast uncertainty. Full-sample spillovers are moderate and selective, with meaningful channels involving S\&P500 turbulence, fiat turbulence, VIX, and network clustering. Rolling connectedness reveals that aggregate spillovers rise sharply during broad stress episodes, especially around the onset of the COVID-19 crisis, and that transmitter--receiver roles vary over time. The threshold VAR provides the clearest evidence of state dependence. Low-turbulence periods are characterised by limited spillovers, with transmission mainly via equity turbulence and volatility expectations. In high-turbulence periods, fiat turbulence becomes the dominant transmitter, while network clustering and modularity play a larger role in propagating forecast uncertainty. At the block level, markets, network structure, and macro-financial conditions become substantially more coupled under stress. These findings support that network structure is emergent and state-dependent. It becomes more involved in stress transmission without acting as a persistent exogenous driver of turbulence.

The results have implications for portfolio allocation, risk monitoring, and macro-financial surveillance. The findings suggest that diversification benefits are regime-dependent. Asset classes that appear segmented in calm periods can become locally clustered and compositionally mixed under stress, reducing the protection normally expected from cross-asset diversification. Risk-management frameworks should therefore monitor not only volatility and pairwise correlations, but also the structural organisation of dependence networks, including clustering, modularity, and cross-class mixing. For macro-financial surveillance, the results indicate that cryptocurrencies are not consistently dominant transmitters of stress, but they are sufficiently integrated to absorb shocks from broader market and currency conditions. The strong role of fiat turbulence in high-turbulence regimes further suggests that currency-market instability can become a central channel of cross-asset propagation when the system is already disturbed.

The analysis is based on correlation networks and reduced-form VAR connectedness, which capture comovement, predictive transmission, and forecast-error variance spillovers, but do not identify structural causal shocks. The results should therefore be interpreted as evidence of state-dependent predictive transmission, not as causal proof that one market structurally drives another. The rolling correlation networks are estimated from short windows relative to the number of assets, so individual pairwise links may contain sampling noise. For this reason, the interpretation focuses on aggregate and consensus-based network summaries rather than individual edges. The balanced sample improves comparability across asset classes, but it necessarily emphasises large and persistent assets and may underrepresent the long tail of cryptocurrencies and equities. In addition, aligning continuously traded cryptocurrencies with traditional market calendars requires interpolating weekend fiat and equity prices, which may attenuate crypto--traditional market correlations during weekend-specific crypto stress episodes.

Future research could compare rolling correlation networks with sparse, partial-correlation, tail-dependence, or high-frequency realised-dependence networks. Structural identification strategies or external instruments could also help distinguish predictive transmission from causal shock propagation. The framework could also be extended to include commodities, bonds, stablecoins, decentralised finance tokens, or cross-border capital-flow indicators. More broadly, the framework could support real-time monitoring of periods in which diversification benefits weaken, and cross-market spillover risk intensifies.

\clearpage

\section*{Acknowledgements}

R.J. and L.E.C.R thank Fabrizio Lillo, Daniele Marinazzo, and Jan Ryckebusch for suggestions. R.J. is funded by the China Scholarship Council (CSC) from the Ministry of Education of P. R. China. L.E.C.R. is partially funded by the Bijzonder Onderzoeksfonds (BOF) (2024/01/709) from Ghent University, Belgium.

\bibliography{maintext_RJLR}

\setcounter{figure}{0}
\setcounter{table}{0}
\setcounter{section}{0} 
\renewcommand{\thesection}{\arabic{section}}  
\renewcommand{\thesubsection}{\arabic{section}.\arabic{subsection}}  
\renewcommand{\thefigure}{S\arabic{figure}}
\renewcommand{\thetable}{S\arabic{table}}

\section*{Supplementary Information}

\section{Asset lists}
\label{sec:DataSI}

Table~\ref{Tab:cryptocurrency} shows the code of the cryptocurrencies used in our study.

\begin{table}[ht]
    \centering
    \small
    \begin{tabular}{llllllllll}
    \hline
        ADA    & ADX   & AE    & AION  & AMB   & ANT   & AOAR  & ARDR  & ARK   & AST   \\
        BAT    & BCD   & BCH   & BCN   & BLOCK & BNB   & BNT   & BTC   & BTM   & BTS   \\
        BTU    & NEBL  & NEO   & NLG   & NMC   & NMR   & NXS   & NXT   & OAX   & OCEANp \\
        OMG    & PAC   & PART  & PAY   & PIVX  & PLR   & PND   & PPC   & CND   & CVC   \\
        DASH   & DCN   & DCR   & DENT  & DGB   & DIME  & DNT   & DOGE  & EDG   & ETC   \\
        ETH    & ETP   & FCT   & FTC   & FUN   & GAME  & GAS   & PPT   & PRE   & PRO   \\
        QRL    & QTUM  & RCN   & RDD   & RDN   & REV   & RLC   & SALT  & SC    & SCRT  \\
        SNC    & SNM   & SNT   & SOUL  & STEEM & STORJ & STRAX & SWFTC & SYNX  & SYS   \\
        GBYTE  & GLM   & GRC   & ICX   & IGNIS & KEY   & KMD   & KNC   & LBC   & LINK  \\
        LRC    & LSK   & LTC   & LUNA  & MAID  & MANA  & MDA   & MDT   & IOTA  & MKR   \\
        MLN    & MONA  & MTH   & MTL   & MYST  & NAS   & TOA   & TRX   & UBQ   & USDT  \\
        VAL    & VERI  & VET   & VIB   & WAVES & WGR   & XCP   & XEM   & XLM   & XRP   \\
        XTZ    & XVG   & ZAP   & ZCL   & ZEC   & ZEN   & ZRX   &       &       &       \\  \hline
    \end{tabular}
    \caption{The list of the $N=127$ cryptocurrencies used in our study.}
    \label{Tab:cryptocurrency}
\end{table}

Table~\ref{Tab:fiat} shows the code of the fiat currencies used in our study.

\begin{table}[ht]
    \centering
    \small
    \begin{tabular}{llllllllll}
    \hline
         AED  &  AFN  &  ALL  &  ANG  &  ARS  &  AUD  &  AWG  &  BBD  &  BDT  &  BGN  \\
         BHD  &  BIF  &  BMD  &  BND  &  BOB  &  BRL  &  BSD  &  BWP  &  BYN  &  BZD  \\
         CAD  &  CDF  &  CHF  &  CLP  &  CNY  &  COP  &  CRC  &  CVE  &  CZK  &  DJF  \\
         DKK  &  DOP  &  DZD  &  EGP  &  ETB  &  EUR  &  FJD  &  GBP  &  GHS  &  GMD  \\
         GNF  &  GTQ  &  GYD  &  HKD  &  HNL  &  HTG  &  HUF  &  IDR  &  ILS  &  INR  \\
         IQD  &  IRR  &  ISK  &  JMD  &  JOD  &  JPY  &  KES  &  KHR  &  KMF  &  KRW  \\
         KWD  &  KYD  &  KZT  &  LAK  &  LKR  &  LRD  &  LSL  &  LYD  &  MAD  &  MDL  \\
         MGA  &  MKD  &  MMK  &  MOP  &  MUR  &  MVR  &  MWK  &  MXN  &  MYR  &  MZN  \\
         NAD  &  NGN  &  NIO  &  NOK  &  NPR  &  NZD  &  OMR  &  PAB  &  PEN  &  PGK  \\
         PHP  &  PKR  &  PLN  &  PYG  &  QAR  &  RON  &  RSD  &  RUB  &  RWF  &  SAR  \\
         SCR  &  SDG  &  SEK  &  SGD  &  SLL  &  SOS  &  SVC  &  SZL  &  THB  &  TMT  \\
         TND  &  TRY  &  TTD  &  TWD  &  TZS  &  UAH  &  UGX  &  UYU  &  UZS  &  VND  \\
         XAF  &  XCD  &  XOF  &  XPF  &  YER  &  ZAR  &  ZMW  &           &           &           \\
    \hline
    \end{tabular}
    \caption{The list of the $N=127$ fiat currencies used in our study.}
    \label{Tab:fiat}
\end{table}

Table~\ref{Tab:SP500} shows the code of the S\&P500 stocks used in our study.

\begin{table}[ht]
    \centering
    \small
    \begin{tabular}{llllllllll}
    \hline
        AVGO & MSFT & AAPL & NVDA & AMZN & META & GOOGL & GOOG & LLY  & TSLA \\
        JPM  & XOM  & WMT  & UNH  & V    & MA   & JNJ   & HD   & PG   & COST \\
        AMD  & MRK  & ORCL & ABBV & CRM  & CVX  & BAC   & NFLX & ADBE & KO   \\
        ACN  & PEP  & TMO  & LIN  & MCD  & ABT  & DIS   & CSCO & TMUS & DHR  \\
        WFC  & INTU & INTC & QCOM & AMAT & IBM  & CMCSA & CAT  & VZ   & NOW  \\
        AXP  & TXN  & UNP  & AMGN & PFE  & ISRG & MS    & LOW  & GE   & PM   \\
        SPGI & SYK  & COP  & HON  & LRCX & GS   & NKE   & BA   & PLD  & BLK  \\
        RTX  & BKNG & T    & SCHW & ETN  & ELV  & NEE   & VRTX & TJX  & PGR  \\
        UPS  & REGN & MDT  & MU   & C    & SBUX & CB    & ADP  & LMT  & DE   \\
        BMY  & MMC  & BSX  & PANW & ADI  & KLAC & MDLZ  & CI   & AMT  & CVS  \\
        BX   & SNPS & ANET & GILD & FI   & ZTS  & CDNS  & EQIX & SHW  & WM   \\
        HCA  & ICE  & CME  & ITW  & CSX  & GD   & CMG   & SO   & TGT  & MAR  \\
        CL   & SLB  & MCO  & DUK  & PH   & AON  & MCK   &     &     &     \\
    \hline
    \end{tabular}
    \caption{The list of the $N=127$ S\&P500 stocks used in our study.}
    \label{Tab:SP500}
\end{table}

\section{Correlation and network descriptive statistics}

\begin{table}[h!]
\sisetup{table-format=-1.2} 
    \centering
    \begin{tabular}{l|rrr|r}
    \toprule
    ~ &   \textbf{Crypto} &  \textbf{ Fiat} & \textbf{S\&P500}  & \textbf{Assets}  \\ \midrule
    \textbf{Crypto} & 0.35 $\left( 0.30 \right)$ & 0.00 $\left( 0.20 \right)$ & 0.07 $\left( 0.21 \right)$  & 127\\ 
    \textbf{Fiat} &  & 0.10 $\left( 0.31 \right)$ & 0.00 $\left( 0.23 \right)$ & 127 \\ 
    \textbf{S\&P500} &  &  & 0.33 $\left( 0.30 \right)$ & 127 \\ 
    \bottomrule
    \end{tabular}
    \caption{Time-average of the mean correlation $\widetilde{\langle\rho\rangle}$ and standard deviation $\sigma$ (in parenthesis) of log returns $r_{i,t}$ between and within asset classes: crypto, fiat, and S\&P500, where \(\langle\rho\rangle_{t} = \frac{1}{\binom{N}{2}}\sum_{i<j}\rho_{ij,t}\).
    The last column indicates the number of assets (nodes) in each asset class.}
    \label{tab:correlation_stats}
\end{table}

Table~\ref{tab:correlation_stats} reports average within- and between-class correlations over the study period. Within-class correlations are higher for cryptocurrencies and S\&P500 equities than for fiat currencies, indicating stronger average synchronisation within the crypto and equity universes and more heterogeneous movements across exchange rates. Between-class correlations are weak on average, with the largest mean cross-class value observed between cryptocurrencies and S\&P500 equities. The near-zero average correlations between crypto--fiat and fiat--equity pairs indicate limited baseline comovement across these classes, although the reported standard deviations show that pairwise correlations vary substantially over time. Thus, weak average cross-class dependence does not preclude temporary increases in comovement during stress episodes.

\begin{figure}[h!]
    \centering
    \includegraphics[width=1\textwidth]{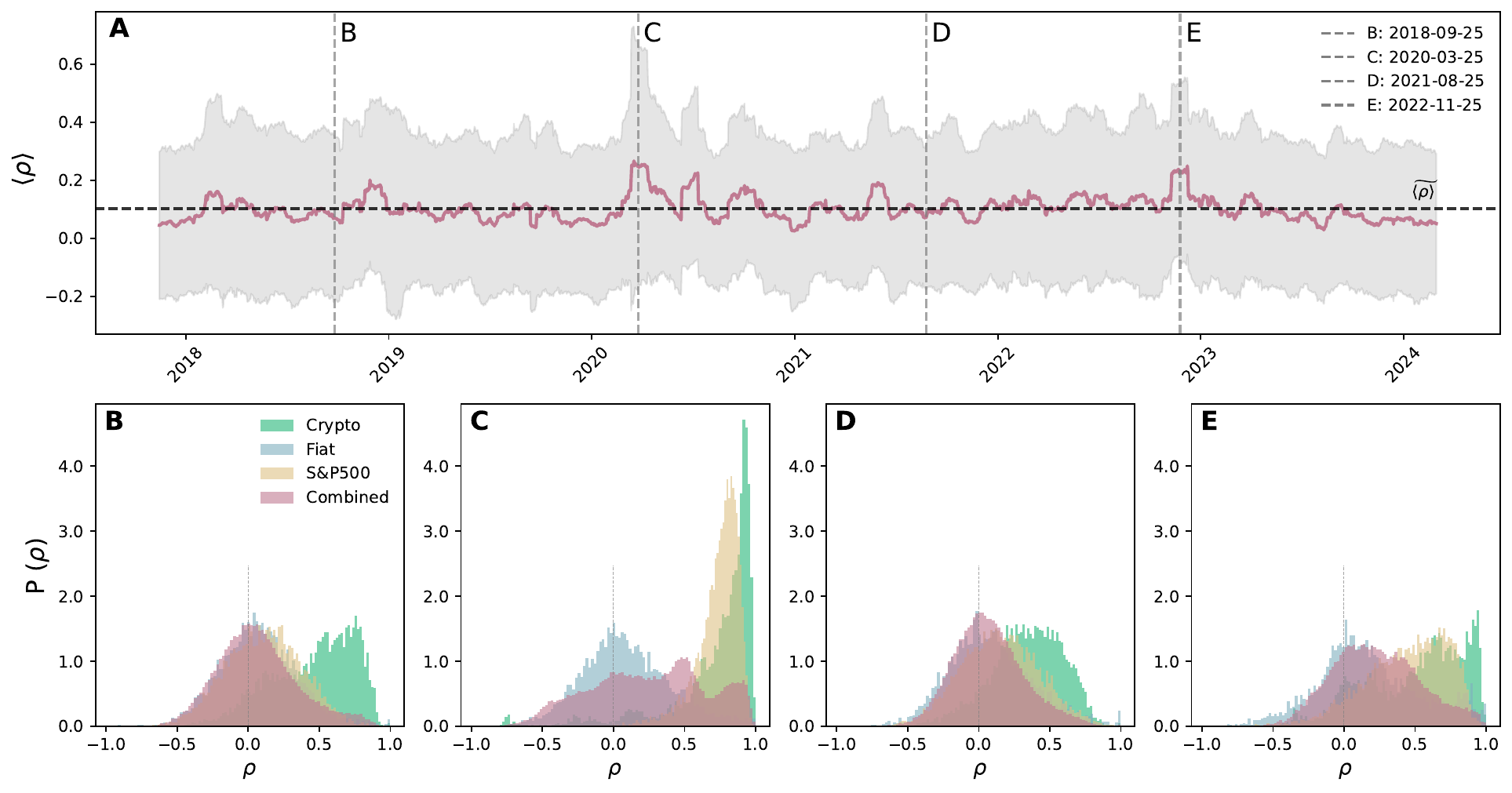}
    \caption{Panel A shows the evolution of the mean correlation $\langle \rho \rangle$ for each time window, with the shaded area indicating the standard deviation. The average of $\langle \rho \rangle$ for the entire study period is \( \widetilde{\langle\rho\rangle} = \frac{1}{T}\sum_{t=1}^{T}{\langle\rho\rangle_{t}} = 0.1\) (horizontal dashed line). 
    The panels (B, C, D, E) show the probability density function P ($\rho$) of correlation $\rho$ values for crypto, fiat, S\&P500, and combined assets, with the dashed line of 0, for specific dates (vertical dashed lines in panel (A)) with and without external shocks. 
    }
    \label{fig:correlation_distributions_and_evolution}
\end{figure}

Figure~\ref{fig:correlation_distributions_and_evolution} illustrates how the distribution of rolling pairwise correlations changes across selected periods. Panel A shows that the mean correlation of the combined system varies over time, with visible increases around major stress episodes. Panels B--E show that the individual asset classes have distinct correlation profiles. Cryptocurrencies and S\&P500 equities are generally shifted toward positive correlations, whereas fiat correlations are more concentrated around zero. During the COVID-19 period, the distributions shift toward stronger positive comovement, especially for S\&P500 equities and the combined asset universe, indicating a broad compression of cross-asset distances. Later periods show a partial moderation of this effect, although the combined distribution remains sensitive to episodes of market stress. These descriptive patterns support the analysis by showing that average correlations conceal substantial temporal variation in the shape of the dependence distribution.

\begin{table}[h!]
\sisetup{table-format=-1.2} 
    \centering
    \begin{tabular}{l|rrrr}
    \toprule
     & \textbf{Crypto} & \textbf{Fiat} & \textbf{S\&P500} & \textbf{Combined}  \\ 
    \midrule
    \textbf{\(N\)} 
        & 127              & 127           & 127          & 381  \\ 
    \textbf{$\widetilde{E}$} 
        & 6778 (656) & 4627 (377) & 6814 (823) & 43904 (3893) \\ 
    \textbf{$\widetilde{\langle\ \kappa \rangle}$}
        & 45.39 (14.51)   & 21.10 (3.71)   & 44.41 (16.93)  & 61.17 (13.39) \\ 
    \textbf{$\widetilde{\langle\ cc \rangle}$}
        & 0.36 (0.12) & 0.18 (0.03) & 0.34 (0.13) & 0.17 (0.04) \\ 
    \bottomrule
    \end{tabular}
    \caption{Characteristics of networks constructed with a correlation threshold of 0, i.e. $\rho_{ij} \leq 0 \rightarrow w_{ij} = 0$, for crypto, fiat, S\&P500, and combined asset classes over the study period. \(N\) is the number of nodes, $\widetilde{E}$ is the time-average of the number of edges, $\widetilde{\langle\ \kappa \rangle}$ is the time-average of the mean weighted degree for each network, $\widetilde{\langle\ cc \rangle}$ is the time-average of the mean weighted clustering coefficient for each network, with standard deviations in parentheses.
    }
    \label{tab:networks_metrix}
\end{table}

Table~\ref{tab:networks_metrix} reports descriptive properties of the positive-correlation networks for each asset class and for the combined system. The crypto and S\&P500 networks have larger average edge counts, weighted degree, and clustering than the fiat network, indicating stronger within-class comovement in digital assets and equities. The fiat network is comparatively weaker and less clustered, consistent with the heterogeneous macroeconomic and policy drivers of exchange-rate movements. The combined network contains many more edges in absolute terms because it includes all within- and between-class pairs, but its average clustering coefficient is lower than that of the crypto and S\&P500 networks. This reflects the fact that cross-class links are more heterogeneous and less locally triangular than within-class links. Accordingly, the combined network captures both within-class cohesion and weaker cross-class connections, which motivates the focus on modularity and cross-class diversity rather than density alone.

\section{Consensus clustering diagnostics}
\label{sec:NC}

\begin{figure}[h!]
    \centering
    \includegraphics[width=1\textwidth]{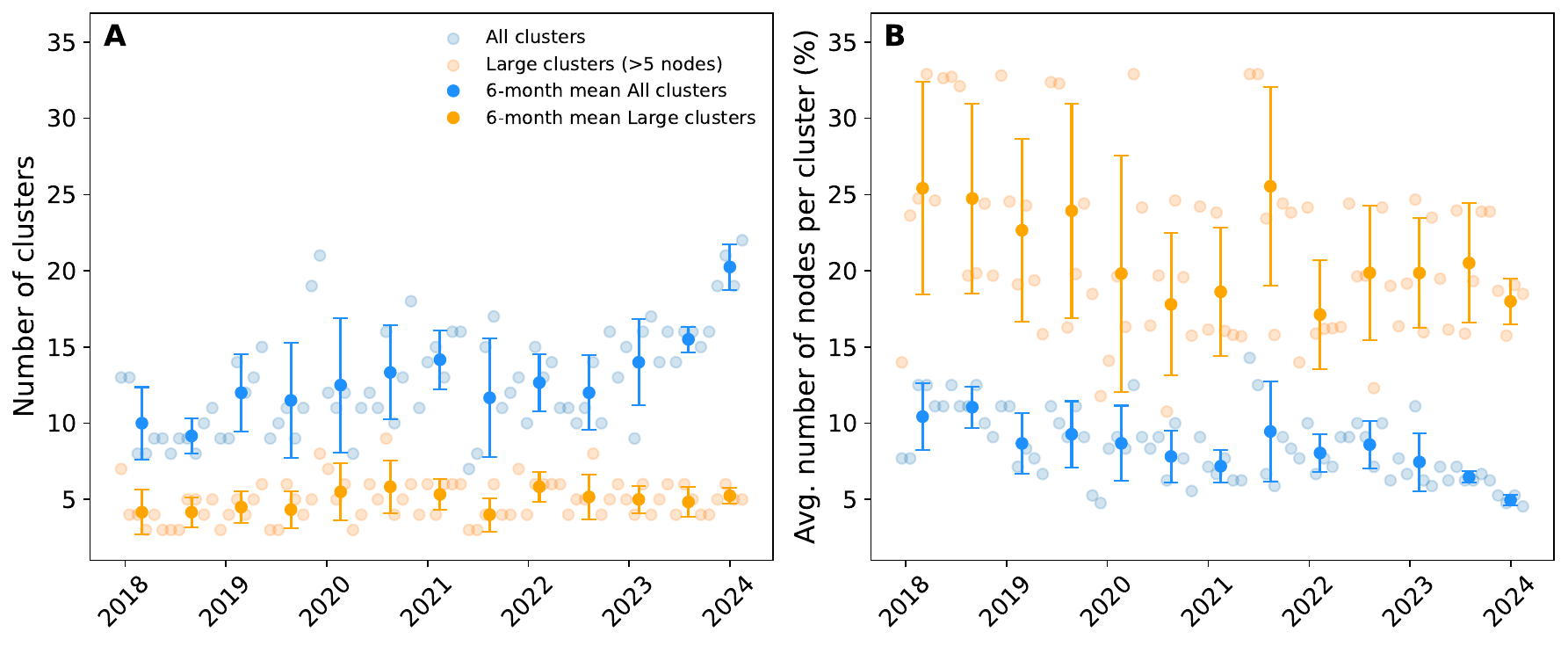}
    \caption{Evolution of robust clusters obtained through the consensus clustering in the combined network. (A) Number of robust clusters over time, where each point represents the consensus clustering derived from every 30 correlation networks with no overlap. 
    Lighter circles represent individual values (for a given $\Delta_c$), dark circles represent 6-month averages, and vertical bars represent the standard deviation; (B) The average fraction of nodes in each cluster over time.}
    \label{fig:robust_clusters}
\end{figure}

To assess whether the modular organisation corresponds to persistent groups rather than single-window partitions, we apply the consensus clustering procedure described in Section~\ref{sec:correlation_networks}. Figure~\ref{fig:robust_clusters}A shows the evolution of the number of stable clusters in the combined network. The number of large clusters is relatively stable, while the total number of stable clusters increases over time, indicating that the network becomes more locally partitioned. Figure~\ref{fig:robust_clusters}B shows a decline in the average fraction of nodes per cluster, particularly after 2022, suggesting that robust clusters become smaller on average.
This cluster-count analysis should be interpreted together with the diversity results in the main text. An increasing number of robust clusters does not necessarily imply stronger asset-class separation. Instead, it indicates greater partitioning of the combined network, while the sufficiently large stable clusters can still become more compositionally mixed across asset classes. Thus, cluster fragmentation and cross-class diversity capture different aspects of network organisation, with the former concerning the number and size of robust groups, whereas the latter concerns their asset-class composition.

\section{VAR specification and diagnostics}

\begin{table}[htbp]
\centering
\small
\setlength{\tabcolsep}{8pt}
\renewcommand{\arraystretch}{1.2}
\begin{tabular}{ccccc}
\toprule
\textbf{Lag} & \textbf{AIC} & \textbf{BIC} & \textbf{FPE} & \textbf{HQIC} \\
\midrule
\textbf{0}  & -0.3944 & -0.3716 & 0.6741  & -0.3860 \\
\textbf{1}  & -1.3150 & -1.0870 & 0.2685  & -1.2320 \\
\textbf{2}  & -1.6270 & -1.1940 & 0.1965  & -1.4690 \\
\textbf{3}  & -1.9250 & -1.2860 & 0.1459  & -1.6920 \\
\textbf{4}  & -2.2410 & -1.3980 & 0.1063  & -1.9330 \\
\textbf{5}  & -2.4760 & -1.4270 & 0.0841  & -2.0930 \\
\textbf{6}  & -2.6940 & \textbf{-1.4400*} & 0.0676  & \textbf{-2.2360*} \\
\textbf{7}  & -2.7200 & -1.2610 & 0.0659  & -2.1870 \\
\textbf{8}  & -2.7360 & -1.0710 & 0.0649  & -2.1280 \\
\textbf{9}  & \textbf{-2.7390*} & -0.8692 & \textbf{0.0647*} & -2.0570 \\
\textbf{10} & -2.7250 & -0.6504 & 0.0656  & -1.9680 \\
\bottomrule
\end{tabular}
\caption{VAR lag order selection criteria with lag orders from 0 to 10. AIC denotes the Akaike Information Criterion, BIC the Bayesian Information Criterion, FPE the Final Prediction Error, and HQIC the Hannan--Quinn Information Criterion. An asterisk (*) indicates the lag length selected by the corresponding criterion.}
\label{tab:lag_selection}
\end{table}

Lag order selection for the VAR model is reported in Table~\ref{tab:lag_selection}. The VAR is estimated on standardised first differences of all variables to ensure stationarity and comparability across series with heterogeneous scales. Information criteria provide mixed guidance, where the AIC and FPE favour a higher lag order ($p=9$), while the more parsimonious BIC and HQIC select $p=6$. Given the dimensionality of the system and the need to preserve degrees of freedom, we adopt VAR(6), which is selected by BIC and HQIC and provides a parsimonious specification relative to the AIC/FPE choice of \(p=9\)~\citep{kilian2006new, kilian2017structural}. The estimated VAR(6) satisfies the stability condition, with all eigenvalues lying strictly inside the unit circle, ensuring a well-defined moving-average representation and valid impulse response analysis. 
Residual diagnostics indicate remaining conditional heteroskedasticity and minor serial dependence in the turbulence indices. These features are common in volatility- and stress-related financial series and do not invalidate the reduced-form VAR analysis, but they motivate the use of robust/bootstrap procedures rather than the homoskedastic Gaussian assumption~\citep{engle2002dynamic, andersen2003modeling, hamilton2020time}. 
Rather than over-fitting the lag structure to eliminate residual dependence, we retain VAR(6) and explicitly address this issue by conducting inference using generalised impulse responses (GIRFs) and generalised forecast error variance decompositions (GFEVD), both complemented with bootstrap confidence bands~\citep{gonccalves2004bootstrapping}. This approach follows established practice in the spillover literature, where emphasis is placed on the patterns, persistence, and relative magnitudes of dynamic transmission rather than on exact pointwise significance~\citep{diebold2014network, barunik2018measuring}.

\section{Full-sample predictive and impulse-response diagnostics}

We use Granger causality~\citep{granger1969investigating} tests to identify directional predictive links between variables. For a pair of variables \(i\rightarrow j\), the test asks whether past values of variable \(i\) improve the prediction of variable \(j\), after controlling for the past values of all other variables in the system. In the VAR, this corresponds to testing whether the coefficients on the lagged values of \(i\) are jointly zero in the equation for \(j\). 
Let \(\Phi_{\ell}(j,i)\) denote the coefficient linking variable \(i\) at lag \(\ell\) to variable \(j\). 
The null hypothesis is
\begin{equation}
H_0: \Phi_{\ell}(j,i)=0
\quad
\text{for all}
\quad
\ell=1,\ldots,p.
\label{eq:granger_null}
\end{equation}
Rejecting this null means that variable \(i\) contains useful past information for predicting variable \(j\). 
These tests are directional and predictive, but they do not quantify the magnitude of spillovers and should not be interpreted as evidence of structural causality.

\begin{table}
\footnotesize
\centering
\begin{tabular}{l|lllllllll}
\toprule
\diagbox{\textbf{Causing}}{\textbf{Caused}}  
& $\boldsymbol{\text{TI}_{\textbf{s\&p500}}}$ 
& $\boldsymbol{\text{TI}_{\textbf{fiat}}}$ 
& $\boldsymbol{\text{TI}_{\textbf{crypto}}}$ 
& \textbf{$\left\langle \boldsymbol{cc} \right\rangle$}   
& \textbf{Q} 
& \textbf{$\left\langle \boldsymbol{D} \right\rangle$} 
& \textbf{VIX} 
& \textbf{USDX} 
& \textbf{EPU} \\
\midrule
$\boldsymbol{\text{TI}_{\textbf{s\&p500}}}$ 
&  & <0.001 & 0.016 & 0.794 & 0.009 & 0.479 & 0.002 & 0.183 & <0.001 \\
$\boldsymbol{\text{TI}_{\textbf{fiat}}}$ 
& 0.013 &  & <0.001 & 0.007 & 0.334 & 0.426 & 0.560 & 0.332 & 0.001 \\
$\boldsymbol{\text{TI}_{\textbf{crypto}}}$ 
& 0.347 & 0.997 &  & 0.800 & 0.058 & 0.001 & 0.629 & 0.973 & 0.639 \\
\textbf{$\left\langle \boldsymbol{cc} \right\rangle$}   
& 0.870 & 0.963 & 0.320 &  & 0.004 & 0.027 & <0.001 & <0.001 & 0.606 \\
\textbf{Q} 
& 0.316 & 0.866 & 0.598 & <0.001 &  & 0.001 & 0.481 & 0.347 & 0.271 \\
\textbf{$\left\langle \boldsymbol{D} \right\rangle$} 
& 0.335 & 0.180 & 0.166 & 0.296 & 0.490 &  & 0.288 & 0.462 & 0.871 \\
\textbf{VIX} 
& 0.015 & 0.137 & 0.635 & <0.001 & 0.456 & 0.526 &  & <0.001 & 0.188 \\
\textbf{USDX} 
& 0.401 & 0.300 & 0.493 & 0.897 & 0.636 & 0.903 & 0.312 &  & 0.093 \\
\textbf{EPU} 
& 0.002 & 0.145 & 0.994 & 0.365 & 0.176 & 0.258 & 0.412 & 0.030 &  \\
\bottomrule
\end{tabular}
\caption{Full-sample Granger causality tests from the estimated VAR. Entries report p-values for the null hypothesis that lagged values of the row variable do not improve the prediction of the column variable, conditional on the other variables in the system. Rows denote causing variables and columns denote caused variables. Lower p-values indicate stronger evidence of direct lagged predictive content.}
\label{tab:gc_full}
\end{table}

Table~\ref{tab:gc_full} reports pairwise Granger causality tests from the full-sample linear VAR. The tests assess direct lagged predictive content, that whether past values of one variable improve the prediction of another variable after conditioning on the remaining variables in the system. 
The results indicate asymmetric predictive transmission among the three market turbulence indices. S\&P500 turbulence Granger-causes both fiat and cryptocurrency turbulence, while fiat turbulence Granger-causes cryptocurrency turbulence. By contrast, cryptocurrency turbulence does not significantly Granger-cause either S\&P500 or fiat turbulence. This pattern is consistent with a hierarchy in the full-sample predictive dynamics, where turbulence in traditional markets and currency markets contains more information for subsequent cryptocurrency turbulence than the reverse.

The tests also suggest that market turbulence helps predict changes in network structure. S\&P500 turbulence significantly predicts modularity \(Q\), fiat turbulence predicts the average clustering coefficient \(\langle cc\rangle\), and cryptocurrency turbulence predicts both modularity \(Q\) and diversity \(\langle D\rangle\). Because these network indicators are constructed from the combined cross-asset network, this pattern supports the interpretation that market-specific turbulence reshapes system-wide network topology in the average full-sample dynamics. Internal predictive links are also present within the network block, especially between \(\langle cc\rangle\), \(Q\), and \(\langle D\rangle\), indicating that local cohesion, modular segmentation, and cross-class mixing co-evolve once changes in market structure occur.

Network indicators exhibit limited direct predictive power for market turbulence in the full-sample linear Granger causality tests. This does not contradict the threshold-VAR results reported in the main text. The Granger causality tests impose a single parameter structure across all observations and therefore summarise average predictive relationships across both low- and high-turbulence states. By contrast, the threshold VAR separates observations by turbulence regime and shows that the role of network structure changes when the system is in an elevated-turbulence state. 
In particular, \(\langle cc\rangle\) changes from a mild net receiver in the low-turbulence regime to a net transmitter in the high-turbulence regime, indicating that local network cohesion becomes more involved in the propagation of forecast uncertainty under stress. Thus, the two analyses answer different questions, that the full-sample Granger tests assess direct average lagged predictability, whereas the regime-specific connectedness measures assess how forecast-uncertainty transmission is reorganised across turbulence states.

Among macro-financial variables, predictive links to market turbulence are concentrated mainly in S\&P500 turbulence. Both VIX and EPU Granger-cause S\&P500 turbulence, while neither significantly predicts fiat or cryptocurrency turbulence at conventional levels. The reverse channels are also informative, where S\&P500 turbulence Granger-causes both VIX and EPU, and fiat turbulence Granger-causes EPU. USDX plays a more limited role in directly predicting market turbulence, but it is linked to EPU, suggesting that dollar conditions and policy uncertainty interact more through the macro-financial block than through direct market-turbulence prediction. 
Overall, the Granger causality results suggest that, in the full-sample linear system, market turbulence variables provide the clearest direct predictive signals, macro-financial variables mainly interact with equity turbulence and policy uncertainty, and network topology behaves primarily as an adaptive market-structure channel rather than as an average exogenous source of turbulence.

\begin{figure}[h!]
    \centering
    \includegraphics[width=0.7\textwidth]{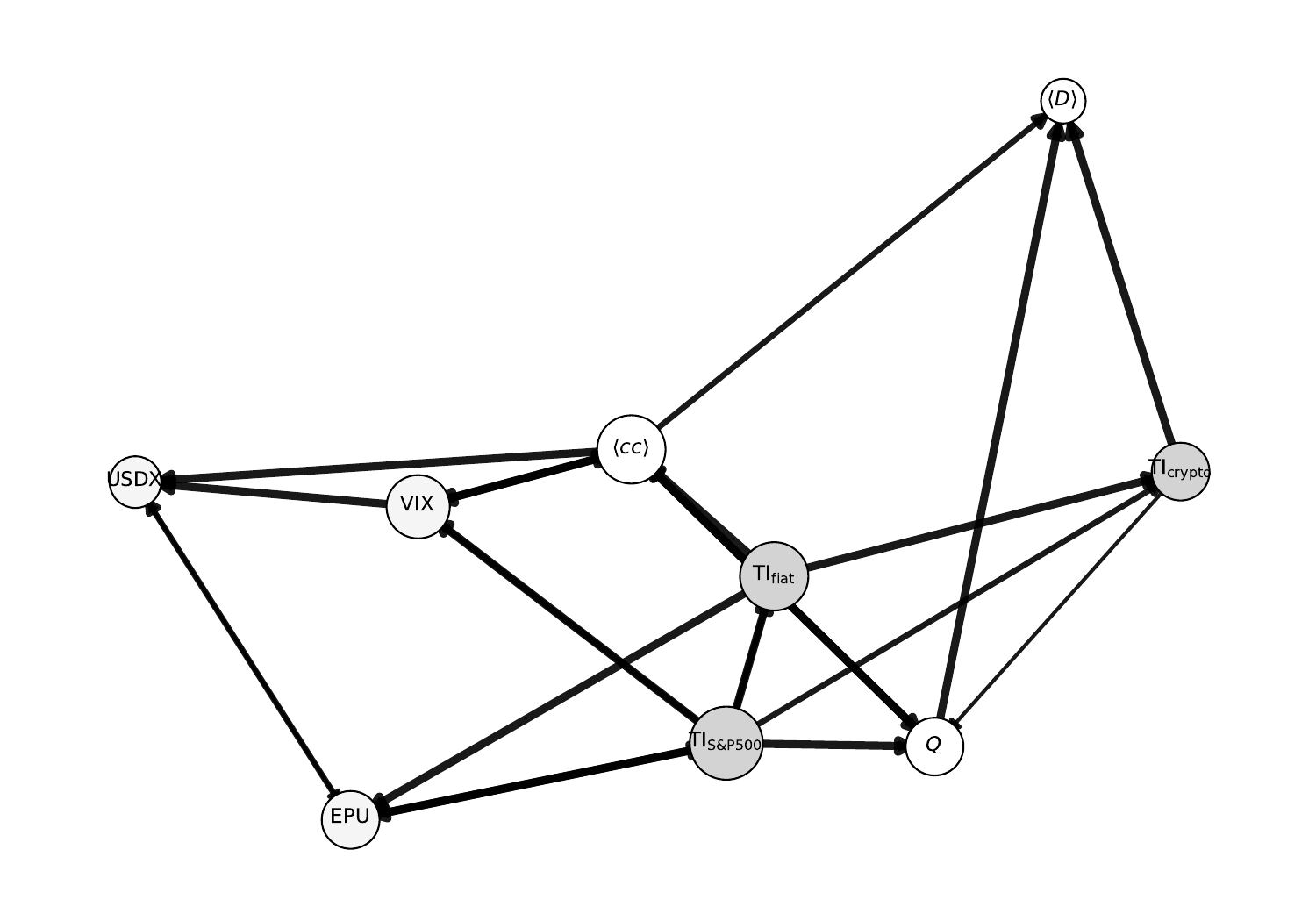}  
    \caption{Directed predictive network based on pairwise Granger causality tests from the estimated VAR. Nodes represent variables and directed edges indicate statistically significant Granger-causal links (\(p<0.1\)). Edge thickness increases with statistical significance. The network layout is based on the Fruchterman--Reingold force-directed algorithm with a fixed random seed for reproducibility.}
    \label{fig:Spillover_Network}
\end{figure}

Figure~\ref{fig:Spillover_Network} provides a network representation of the full-sample Granger causality results. The plot should be interpreted as a map of direct average lagged predictive links in the linear VAR, not as a structural causal graph. Within this full-sample predictive system, market turbulence variables occupy a relatively upstream position, with S\&P500 turbulence showing several outgoing links to fiat turbulence, cryptocurrency turbulence, macro-financial variables, and network structure. Fiat turbulence acts as an intermediate transmitter, while cryptocurrency turbulence has fewer outgoing predictive links to the traditional markets. 
The network indicators appear mainly as receivers in the full-sample Granger graph, consistent with the interpretation that market turbulence reshapes the system-wide dependence topology on average. However, this does not imply that network topology is irrelevant for transmission. As shown in the threshold-VAR results in the main text, the role of network structure is state-dependent, which is weakly predictive in the pooled linear system but becomes more involved in forecast-uncertainty propagation under high-turbulence states. The Granger network therefore complements, rather than substitutes for, the regime-dependent connectedness analysis.

\begin{figure}[h!]
    \centering
    \includegraphics[width=1\textwidth]{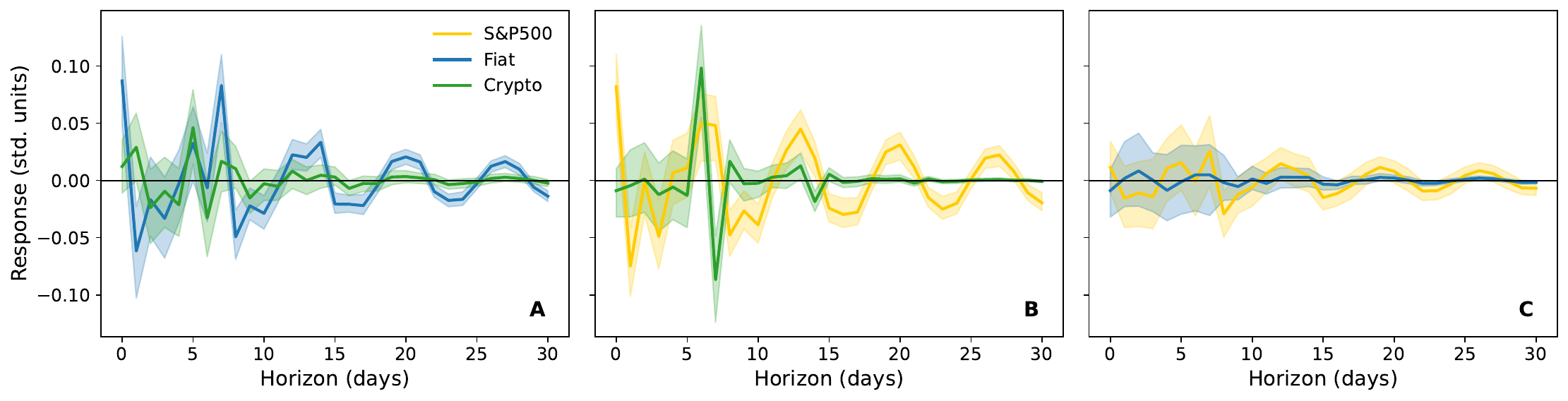}
    \includegraphics[width=1\textwidth]{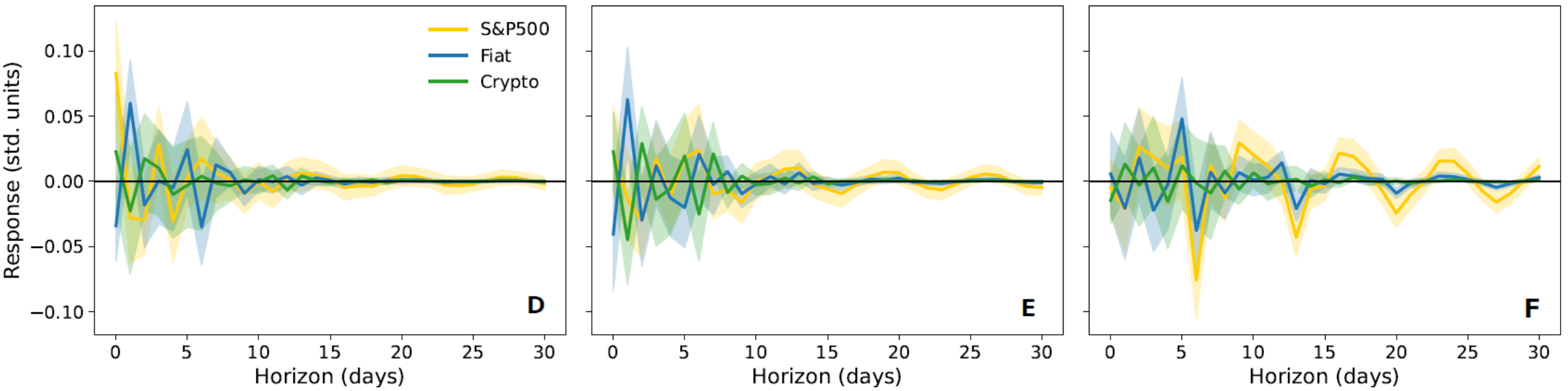} 
    \includegraphics[width=1\textwidth]{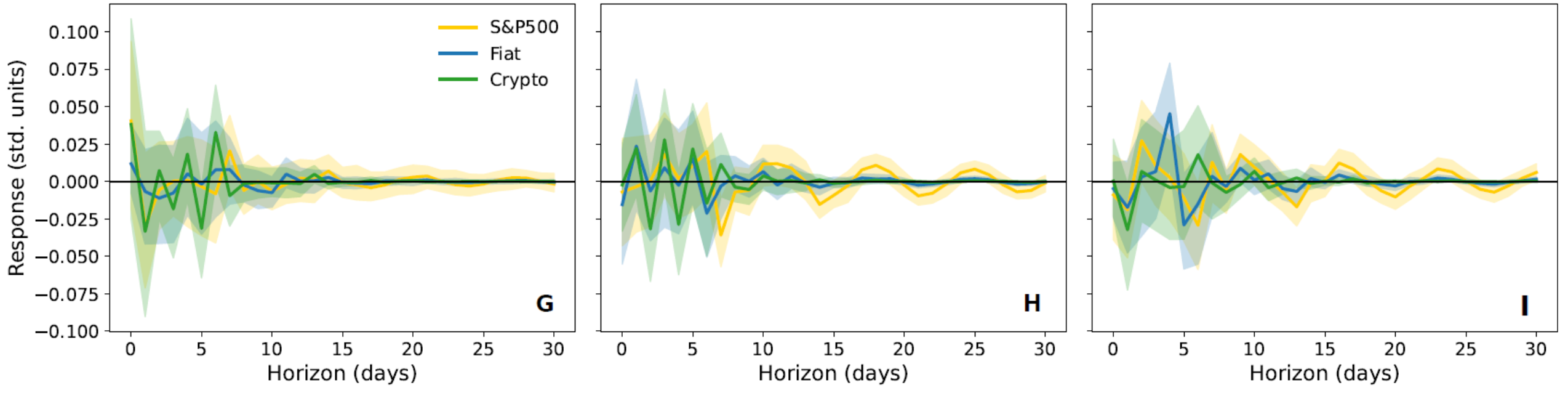}
    \caption{Supplementary full-sample generalised impulse response functions from the estimated VAR. Panels A--C report cross-market responses among the S\&P500, fiat, and cryptocurrency turbulence indices following shocks to (A) S\&P500 turbulence, (B) fiat turbulence, and (C) cryptocurrency turbulence. Panels D--F report responses of market turbulence to macro-financial shocks in (D) VIX, (E) USDX, and (F) EPU. Panels G--I report responses of market turbulence to shocks in network structure, namely (G) \(\langle cc\rangle\), (H) \(Q\), and (I) \(\langle D\rangle\). Responses are computed using the Pesaran--Shin generalised framework, are expressed in standard-deviation units, and are traced over a 30-day horizon. Shaded areas denote 90\% bootstrap confidence intervals.}
    \label{fig:girfs_markets}
\end{figure}

Figure~\ref{fig:girfs_markets} reports supplementary full-sample GIRFs that complement the Granger-causality diagnostics by showing the dynamic response paths following one-standard-deviation shocks.
Panels A--C show cross-market responses among the three turbulence indices. Consistent with the connectedness and Granger-causality results, shocks to S\&P500 turbulence generate short-run responses in fiat and cryptocurrency turbulence, with the stronger response appearing in fiat turbulence. Shocks to fiat turbulence also affect the other markets, including a delayed response in cryptocurrency turbulence. By contrast, shocks originating in cryptocurrency turbulence generate weaker responses in S\&P500 and fiat turbulence, supporting the interpretation that cryptocurrency turbulence is more reactive than transmissive in the full-sample linear system.
Panels D--F show the responses of market turbulence to macro-financial shocks. The responses are concentrated at short horizons and generally converge back toward zero, indicating that macro-financial shocks mainly generate transitory adjustments rather than persistent displacement in market turbulence. A VIX shock in Panel D produces immediate responses across the three markets, with particularly visible early movements in S\&P500 and fiat turbulence. This suggests that changes in implied equity volatility are associated not only with equity-market stress but also with short-run adjustments in currency and cryptocurrency turbulence through the broader risk environment. The response to a USDX shock in Panel E is more oscillatory, especially for fiat turbulence, which is consistent with the direct relevance of broad dollar movements for currency-market stress. S\&P500 and cryptocurrency turbulence respond more weakly and less persistently. Panel F shows that EPU shocks generate short-run fluctuations in market turbulence, with the clearest oscillatory response appearing in S\&P500 turbulence, while fiat and cryptocurrency responses are smaller and surrounded by wider confidence bands. 
Panels G--I show the responses of market turbulence to shocks in network structure. These responses are generally smaller and less persistent than the market-to-market and macro-to-market responses. This supports the main-text interpretation that, in the full-sample linear VAR, network topology is not an average exogenous driver of market turbulence. However, this does not rule out a state-dependent network channel. The threshold-VAR results show that network-related transmission becomes more pronounced under high-turbulence states, meaning that the weak full-sample response partly reflects averaging over calm and stressed periods.

\section{Threshold VAR robustness and regime diagnostics}
\label{sec: robustness_tvar}

\begin{table}
\centering
\begin{tabular}{lr}
\toprule
 & Value \\
\midrule
Threshold variable & TI level \\
Delay $d$ & 1 \\
Lag order $p$ & 4 \\
Criterion & BIC \\
$\hat{\gamma}$ & 518.2384 \\
Low regime obs & 1510 \\
High regime obs & 755 \\
Low share & 0.6667 \\
High share & 0.3333 \\
Quantile of $\hat{\gamma}$ & 0.6667 \\
Regime switches & 708 \\
\bottomrule
\end{tabular}
\caption{Threshold selection and regime composition for the two-regime TVAR. The threshold \(\hat{\gamma}\) is estimated by grid search using the Bayesian Information Criterion (BIC) and applied to the turbulence index in levels \(TI_{t-d}\) with delay \(d=1\). The table reports regime sample sizes, regime shares, and the number of regime switches in the resulting state sequence.}
\label{tab:tvar_threshold}
\end{table}

Table~\ref{tab:tvar_threshold} summarises the data-driven regime definition obtained from BIC-based threshold selection. 
Because the threshold VAR estimates separate parameter sets in each regime, we re-selected the lag order for the TVAR using BIC within the threshold-search procedure, which selected \(p=4\). The estimated threshold assigns two-thirds of the sample to the low-turbulence regime and one-third to the high-turbulence regime. The resulting regime sequence contains 708 switches, indicating that the high-turbulence state should not be interpreted as a small number of prolonged crisis episodes. Rather, it identifies days on which the joint cross-asset return configuration is unusually displaced relative to its recent multivariate benchmark. Accordingly, the regime-conditional GIRFs describe shock propagation conditional on an elevated-displacement state, not the response during a single sustained historical episode.

\begin{table}
\centering
\begin{tabular}{lrr}
\toprule
Regime & max $|\lambda|$ & Stable (<1) \\
\midrule
Low & 0.860 & True \\
High & 0.946 & True \\
\bottomrule
\end{tabular}
\caption{Stability diagnostics for regime-specific VAR dynamics. The table reports the maximum modulus of the companion-matrix eigenvalues for the low- and high-turbulence regimes. Values below one indicate that each regime VAR is stable (covariance-stationary), ensuring that impulse responses and variance decompositions are well-defined.}
\label{tab:tvar_stability}
\end{table}

Table~\ref{tab:tvar_stability} confirms that both regime-specific VAR dynamics satisfy the stability condition (maximum eigenvalue modulus below unity), implying that the regime-conditional impulse responses and FEVD-based connectedness measures are well-defined. 

\begin{table}
\centering
\begin{tabular}{lrrr}
\toprule
H & TSI Low & TSI High & Ratio (High/Low) \\
\midrule
30 & 8.617 & 24.345 & 2.825 \\
60 & 8.618 & 24.398 & 2.831 \\
90 & 8.618 & 24.399 & 2.831 \\
\bottomrule
\end{tabular}
\caption{Horizon robustness of regime-dependent connectedness. Total Spillover Index (TSI) estimates are reported for forecast horizons \(H\in\{30,60,90\}\) days under each regime. The ratio column summarises the multiplicative difference in system-wide connectedness between high- and low-turbulence regimes.}
\label{tab:tvar_tsi}
\end{table}

Table~\ref{tab:tvar_tsi} shows that the regime contrast in total connectedness is robust to the forecast horizon across $H=30-90$ days, the high-turbulence regime exhibits consistently higher spillovers (approximately three times the low-regime level), indicating that the state dependence is not driven by horizon choice.

\begin{figure}[h!]
    \centering
    \includegraphics[width=1\textwidth]{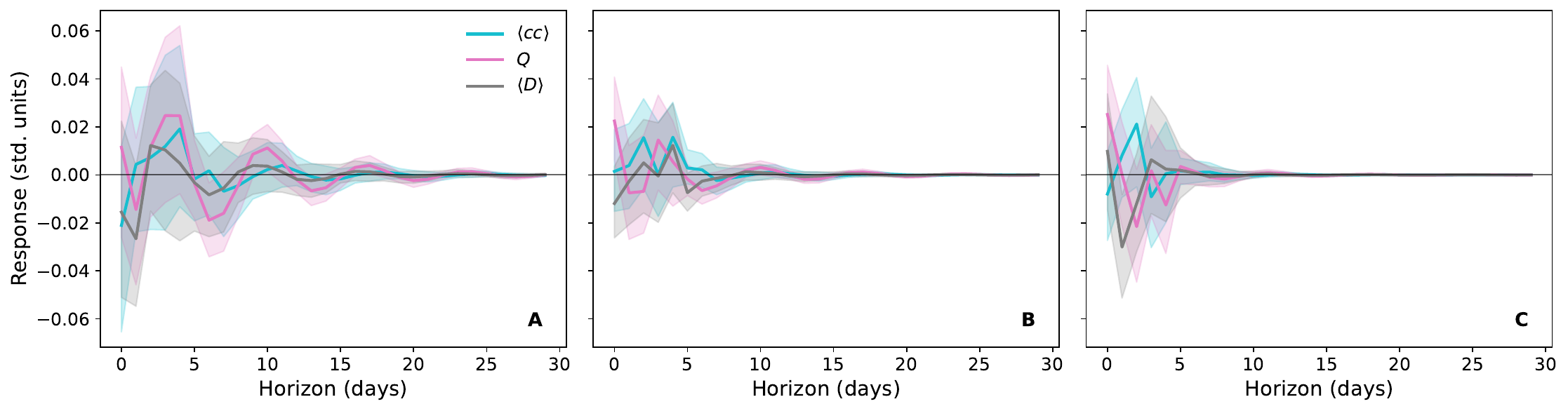} 
    \includegraphics[width=0.97\textwidth]{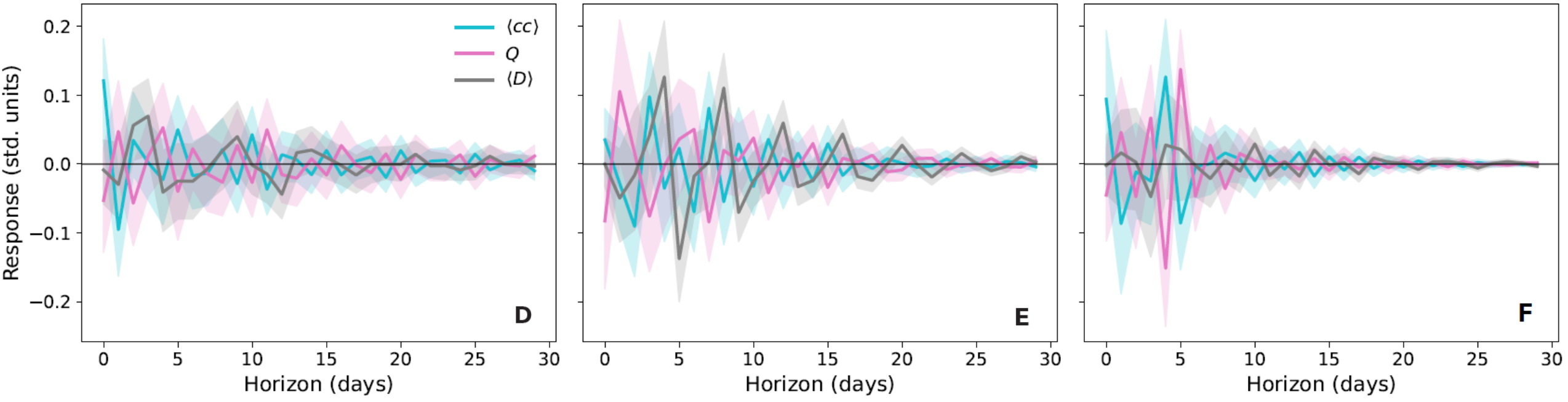}  
    \caption{Regime-conditional generalised impulse response functions (GIRFs) from network topology shocks to market turbulence under the threshold VAR. Panels A--C report the responses of (A) \(\mathrm{TI}_{\mathrm{S\&P500}}\), (B) \(\mathrm{TI}_{\mathrm{fiat}}\), and (C) \(\mathrm{TI}_{\mathrm{crypto}}\) to one-standard-deviation shocks in network clustering \(\langle cc\rangle\), modularity \(Q\), and diversity \(\langle D\rangle\) in the low-turbulence regime. Panels D--F report the corresponding responses of (D) \(\mathrm{TI}_{\mathrm{S\&P500}}\), (E) \(\mathrm{TI}_{\mathrm{fiat}}\), and (F) \(\mathrm{TI}_{\mathrm{crypto}}\) in the high-turbulence regime. All responses are computed over a horizon of \(H=30\) days. Shaded areas denote bootstrap confidence bands, and responses are expressed in standardised units.}
    \label{fig:GIRF_Low_High}
\end{figure}

Figure~\ref{fig:GIRF_Low_High} reports the complementary regime-conditional GIRFs from network topology shocks to market turbulence. In the low-turbulence regime, shocks to \(\langle cc\rangle\), \(Q\), and \(\langle D\rangle\) generate small and short-lived responses in all three market turbulence indices. This is consistent with the main-text result that network measures play a limited initiating role under tranquil conditions.
In the high-turbulence regime, the same topology shocks generate larger and more volatile short-run responses, especially for fiat and cryptocurrency turbulence. The responses remain transitory, but their larger magnitude indicates that, once the system is already in an elevated-displacement state, changes in network topology contain more information about subsequent market-turbulence dynamics. This supports a feedback interpretation, where market stress reshapes network structure, and under high turbulence, the reconfigured topology becomes more involved in short-run propagation. These results should be interpreted as reduced-form evidence of state-dependent transmission, not as structural proof that topology independently causes turbulence.

\section{Robustness to correlation-distance transformation}

\begin{figure}[h!]
    \centering
    \includegraphics[width=1\textwidth]{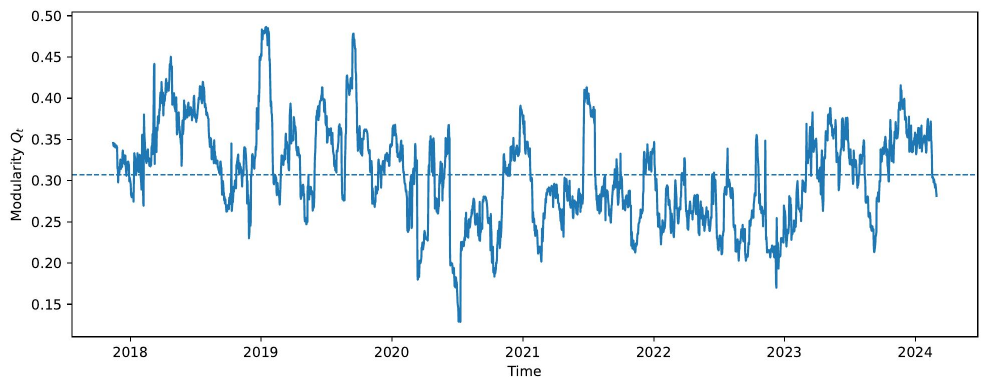}
    \caption{Modularity evolution under the correlation-distance network representation. The solid line shows the modularity \(Q_t\) of the largest connected component over time, and the dashed horizontal line indicates the time average of the modularity series.}
    \label{fig:dist_modularity}
\end{figure}

The baseline network measures are computed from the positive-correlation weighted adjacency matrix. As a robustness check, we recompute the modularity series using the correlation-distance transformation \(d_{ij,t}=\sqrt{2(1-\rho_{ij,t})}\), retaining links with \(d_{ij,t}<\sqrt{2}\), which is equivalent to \(\rho_{ij,t}>0\). Because weighted modularity treats larger weights as stronger connections, distances are converted into proximity weights before computing modularity. 
As shown in Fig.~\ref{fig:dist_modularity}, the distance-transformed modularity series differs in level from the baseline correlation-weighted series but displays a broadly similar temporal pattern. This supports the interpretation that the main modularity dynamics are not driven solely by the choice between correlation weights and distance-based proximity weights.

\end{document}